\documentclass{aa} %

\usepackage{amsmath}
\usepackage{graphicx}
\usepackage{color}

\usepackage{txfonts}
\usepackage{epstopdf} 
\usepackage{siunitx}
\newcommand\gaia{\textit{Gaia}}

\newcommand\gdrtwo{\gaia~DR2}

\newcommand\secref[1]{Sect.~\ref{#1}}

\newcommand\figref[1]{Fig.~\ref{#1}}
\newcommand\figsref[1]{Figs.~\ref{#1}}

\newcommand\equref[1]{Eq.~\eqref{#1}}

\begin{document} 

   \title{\gaia~Data Release~2}
   \titlerunning{\gdrtwo\  . Calibration and mitigation of electronic offsets} 

   \subtitle{Calibration and mitigation of electronic offset effects in the data}

\author{N.C.~Hambly\inst{\ref{inst:edinb}} %
\and M.~Cropper\inst{\ref{inst:mssl}}
\and S.~Boudreault\inst{\ref{inst:mssl},\ref{inst:mpis}}
\and C.~Crowley\inst{\ref{inst:he-spaceesac}} %
\and R.~Kohley\inst{\ref{inst:esac}} %
\and J.H.J.~de~Bruijne\inst{\ref{inst:estec}}
\and C.~Dolding\inst{\ref{inst:mssl}}
\and C.~Fabricius\inst{\ref{inst:0012}} %
\and G.~Seabroke\inst{\ref{inst:mssl}} 
\and M.~Davidson\inst{\ref{inst:edinb}} %
\and N.~Rowell\inst{\ref{inst:edinb}} %
\and R.~Collins\inst{\ref{inst:edinb}} %
\and N.~Cross\inst{\ref{inst:edinb}} %
\and J.~Mart\'in-Fleitas\inst{\ref{inst:auroraesac}}
\and S.~Baker\inst{\ref{inst:mssl}} 
\and M.~Smith\inst{\ref{inst:mssl}} 
\and P.Sartoretti\inst{\ref{inst:0005}}
\and O.~Marchal\inst{\ref{inst:0005}}
\and D.~Katz\inst{\ref{inst:0005}}
\and F.~De Angeli\inst{\ref{inst:0011}}
\and G.~Busso\inst{\ref{inst:0011}}
\and M.~Riello\inst{\ref{inst:0011}}
\and C.~Allende~Prieto\inst{\ref{inst:mssl},\ref{inst:0541},\ref{inst:0542}}
\and S.~Els\inst{\ref{inst:0067},\ref{inst:0007}}
\and L.~Corcione\inst{\ref{inst:0035}}
\and E.~Masana\inst{\ref{inst:0012}}
\and X.~Luri\inst{\ref{inst:0012}} 
\and F.~Chassat\inst{\ref{inst:ads}}
\and F.~Fusero\inst{\ref{inst:ads}}
\and J.F.~Pasquier\inst{\ref{inst:ads}}
\and C.~V\'{e}tel\inst{\ref{inst:ads}}
\and G.~Sarri\inst{\ref{inst:esasd}}
\and P.~Gare\inst{\ref{inst:esasd}}
}

\institute{
Institute for Astronomy, School of Physics and Astronomy, University of Edinburgh, 
Royal Observatory, Blackford Hill, Edinburgh, EH9~3HJ, UK\\ \email{nch@roe.ac.uk}
\label{inst:edinb}
\and
Mullard Space Science Laboratory, University College London, Holmbury St Mary, Dorking, Surrey RH5 6NT, UK 
\label{inst:mssl}
\and
Max Planck Institut für Sonnensystemforschung, Justus-von-Liebig-Weg 3, 37077 G\"{o}ttingen, Germany
\label{inst:mpis}
\and
HE Space Operations BV for ESA/ESAC, Camino Bajo del Castillo s/n, 28691 Villanueva de la Ca{\~n}ada, Spain
\label{inst:he-spaceesac}
\and
ESA, European Space Astronomy Centre (ESAC), Camino Bajo del Castillo s/n, 28691 Villanueva de la Ca{\~n}ada, Spain 
\label{inst:esac}
\and
ESA, European Space Research and Technology Centre (ESTEC), Keplerlaan 1, 2201 AG,  Noordwijk, The Netherlands
\label{inst:estec}
\and Institut de Ci\`{e}ncies del Cosmos, Universitat  de  Barcelona  (IEEC-UB), Mart\'{i}  Franqu\`{e}s  1, E-08028 Barcelona, Spain\relax                                                                  
\label{inst:0012}
\and
Aurora Technology B.V.~for ESA/ESAC, Camino Bajo del Castillo s/n, 28691
Villanueva de la Ca{\~n}ada, Spain
\label{inst:auroraesac}
\and
GEPI, Observatoire de Paris, PSL Research University, CNRS, Univ. Paris Diderot, Sorbonne Paris Cit{\'e}, 5 Place Jules Janssen, 92190 Meudon, France\relax                                             
\label{inst:0005}
\and Institute of Astronomy, University of Cambridge, Madingley Road, Cambridge CB3 0HA, United Kingdom\relax                                                                                                
\label{inst:0011}
\and Instituto de Astrof\'{\i}sica de Canarias, E-38205 La Laguna, Tenerife, Islas Canarias, Spain\relax                                                                                                                     
\label{inst:0541}
\and Universidad de La Laguna, Departamento de Astrof\'{\i}sica, E-38206 La Laguna, Tenerife, Islas Canarias, Spain\relax                                                                                                    
\label{inst:0542}
\and Gaia DPAC Project Office, ESAC, Camino bajo del Castillo, s/n, Urbanizacion Villafranca del Castillo, Villanueva de la Ca\~{n}ada, E-28692 Madrid, Spain\relax                                          
\label{inst:0067}
\and Astronomisches Rechen-Institut, Zentrum f\"{ u}r Astronomie der Universit\"{ a}t Heidelberg, M\"{ o}nchhofstr. 12-14, D-69120 Heidelberg, Germany\relax                                                 
\label{inst:0007}
\and INAF - Osservatorio Astrofisico di Torino, via Osservatorio 20, 10025 Pino Torinese (TO), Italy\relax                                                                                                   
\label{inst:0035}
\and Airbus Defence and Space SAS, 31 Rue des Cosmonautes, 31402 Toulouse Cedex 4, France
\label{inst:ads}
\and Directorate of Science, European Space Research and Technology Centre (ESA/ESTEC), Keplerlaan 1, 2201AZ Noordwijk, The Netherlands
\label{inst:esasd}
}

   \date{ }

\abstract{
{\em Context}. The European Space Agency’s \gaia\ satellite was launched into orbit around L2 in December 2013. This 
ambitious mission has strict requirements on residual systematic errors resulting from instrumental 
corrections in order to meet a design goal of sub-10 microarcsecond astrometry. During the design and build phase
of the science instruments, various critical calibrations were studied in detail to ensure that this goal could be met
in orbit. In particular, it was determined that the video-chain offsets on the analogue side of the analogue--to--digital
conversion electronics exhibited instabilities that could not be mitigated fully by modifications to the flight
hardware.
\\{\em Aims}. We provide a detailed description of the behaviour of the electronic offset levels on short ($<<$1~ms) 
timescales, identifying various systematic effects that are known collectively as "offset non-uniformities". 
The effects manifest themselves as transient perturbations on the gross zero-point electronic offset level that 
is routinely monitored as part of the overall calibration process. 
\\{\em Methods}. Using in-orbit special calibration sequences 
along with simple parametric models, we show how the effects can be calibrated, and how these calibrations are applied 
to the science data. While the calibration part of the process is relatively straightforward, the application of the 
calibrations during science data processing requires a detailed on-ground reconstruction of the readout timing of 
each charge-coupled device (CCD) sample on each device in order to predict correctly the highly time-dependent nature of the corrections. 
\\{\em Results}. We demonstrate the effectiveness of our offset non-uniformity models in mitigating the effects in 
\gaia\ data.
\\{\em Conclusions}. We demonstrate for all CCDs and operating instrument/modes on board \gaia\ that the video-chain 
noise-limited performance is recovered in the vast majority of science samples. 
}


   \maketitle
%


\clearpage
\section{Introduction}
\label{sec:intro}

The European Space Agency `Horizon 2000+' mission \gaia\ was launched in December~2013~\citep{2016A&A...595A...1G}.
This ambitious all--sky astrometric and spectrophotometric survey aims to catalogue approximately one billion
astrophysical sources (primarily stars to V~$\sim20$) over the course of a mission of at least five years.
At the same time, \gaia’s Radial Velocity Spectrometer (RVS) is obtaining medium--resolution spectra of 
the brightest ~150 million sources to V~$\sim15$.
\gaia\ employs the Hipparcos measurement principle~\citep{2011EAS....45..109L}. This consists of the 
superposition of two fields of view, separated by a large angle as afforded by space--based observation,
on the same imaging focal plane with a continuous scanning motion. \gaia\ employs sensitive optical charge--coupled
devices (CCDs) as opposed to the photomultiplier tubes used in the precursor Hipparcos mission.
In order to reach mission goals, the requirements on calibration of the CCD measurements from \gaia\ are extreme.
For example, the requirement is to achieve pure photon noise limited performance in bright source 
location in the imaging
data, that is to say,~to approach the Cram\'{e}r--Rao lower bound. This corresponds to centroiding precision at the level of~$<10^{-2}$~pixels
(e.g.~\citealt{2013PASP..125..580M}) and requires other noise sources, in particular read noise, to
make no significant contribution in the sample error budget.
Instrumental corrections are thus required to leave no residual sample--to--sample fluctuations higher than 
those expected from read noise alone in
situations where that limiting performance should be observed (e.g.~for zero photoelectric signal).
Optical, photometric, and geometric calibrations of the instruments along with the spacecraft attitude must be 
derived primarily from the survey science data~\citep{2016A&A...595A...3F},
and these calibrations are entangled in complex and subtle ways. While the electronic zero-point offset on the
CCD amplification stage (commonly known as the bias level) is easily separable from nearly all other calibrations,
the complexity of the \gaia\ CCD design and operation leads to a quasi--stable behaviour that in turn complicates
what would otherwise be a relatively straightforward additive correction to the data at the head of the
processing chain.

\begin{figure*}
\centerline{\includegraphics[scale=0.25,clip,trim=50 100 150 200]{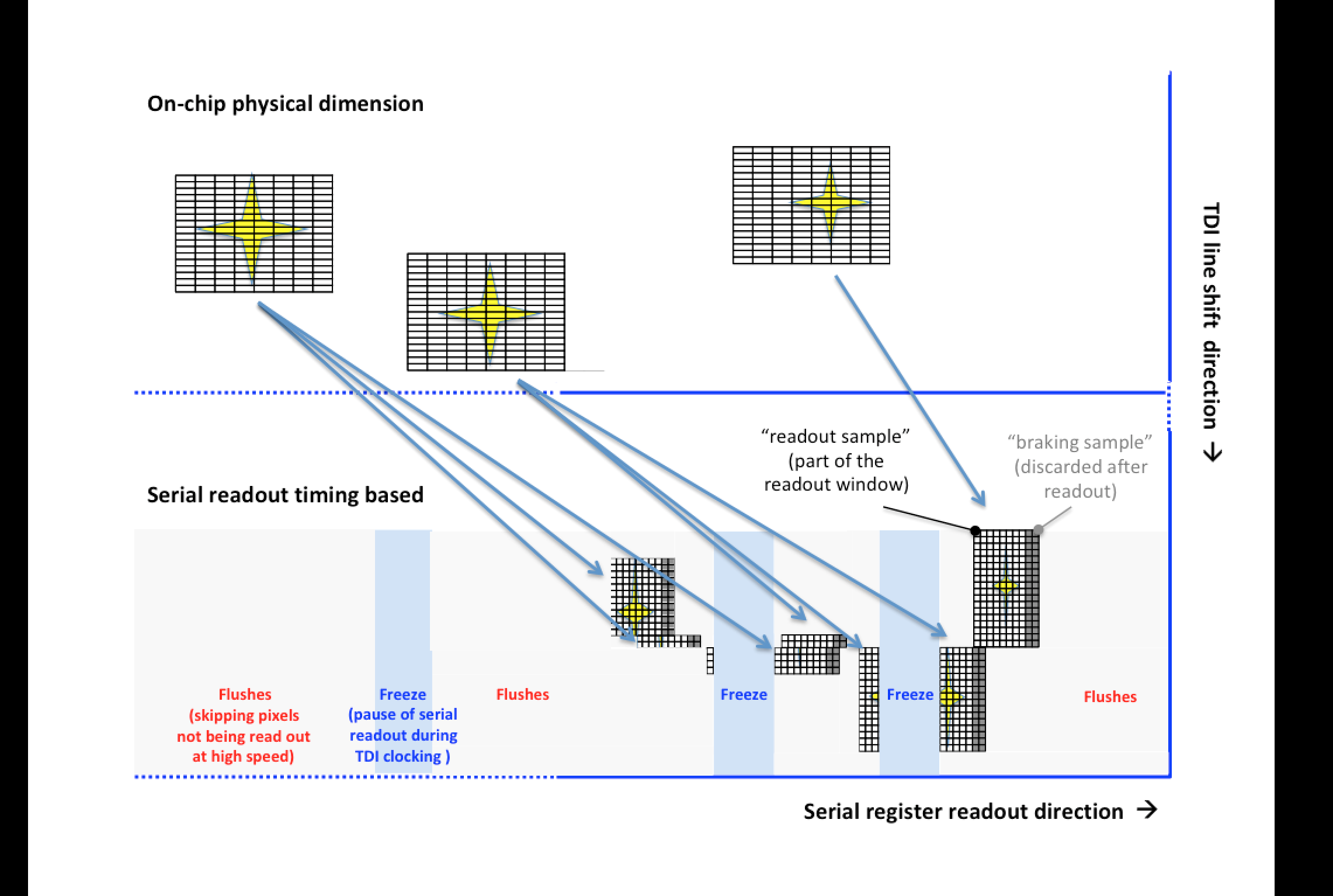}}
\caption{Illustrating the principles of \gaia\ windowed readout: Images and spectra are measured via
windowed sampling with unwanted pixels in between windows being flushed at high speed. The timing of
window samples changes with respect to readout freezes (associated with parallel clock pulses), and
braking samples are inserted after flush sequences where time allows.}
\label{fig:gaiaReadoutIllustration}
\end{figure*}

\begin{figure}
\centerline{\includegraphics[scale=0.45]{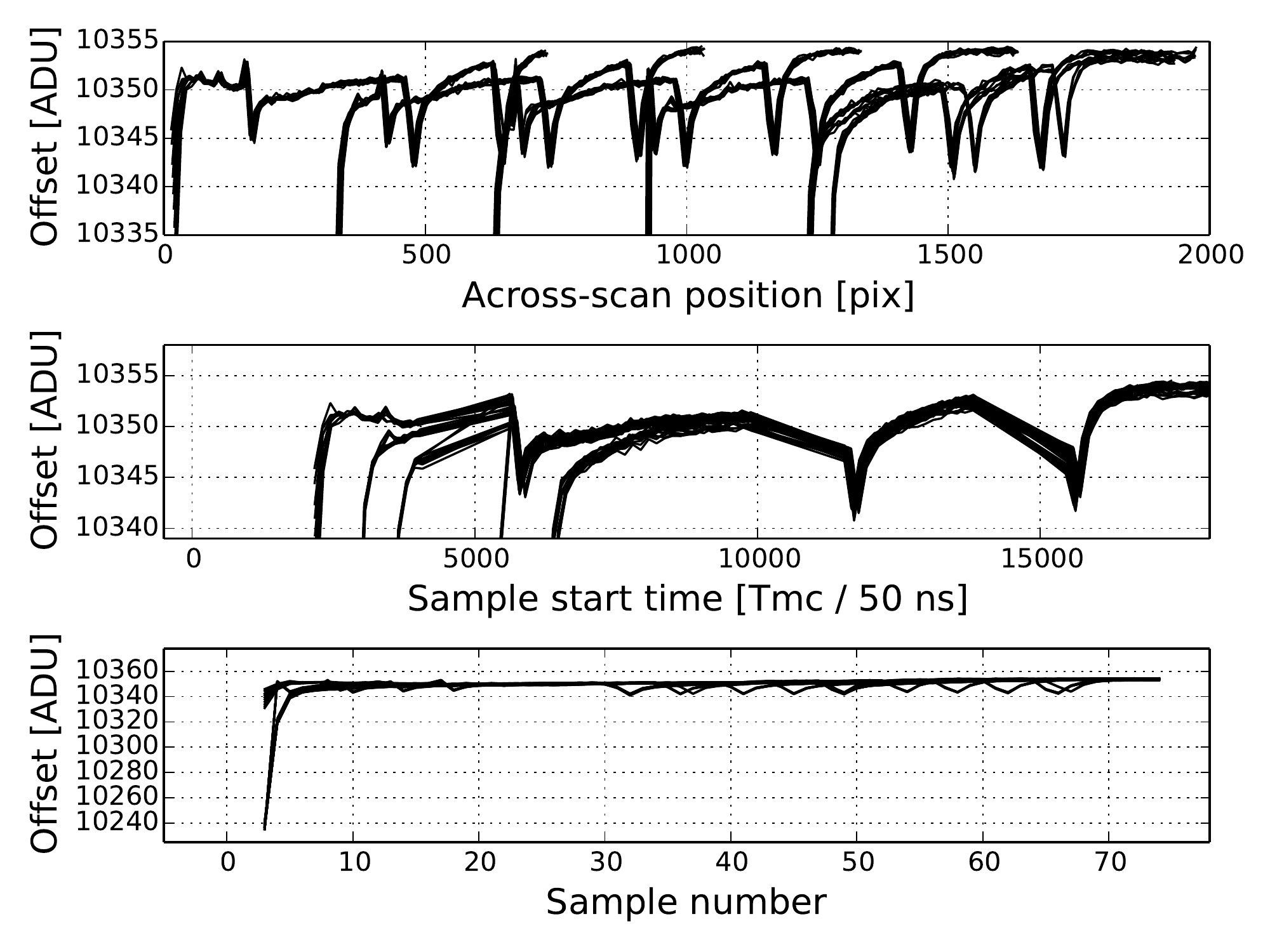}}
\caption{(Top) Change in the electronic bias in digitised units (ADU) as a function of pixel position 
in the serial register for two different window patterns as measured in on--ground pre--flight testing.
Negative excursions with steep declines and slower recoveries are evident: these are associated 
with the flushes. (Centre) Pattern shown in time (units of `master clock' cycles of 50~ns,
denoted Tmc), covering the duration of the readout 
of the serial register. This synchronises the pattern of glitches, as the timing of the pixel phases is 
unvarying. (Bottom) Excursions shown synchronised to the start of each window, for different patterns 
of flush and readout. The excursions are larger in the flushes (lower) than in the glitches (middle).}
\label{fig:test_data}
\end{figure}

The design and operation of the \gaia\ CCDs and associated Proximity Electronics Modules (PEMs) is discussed
elsewhere~\citep{2009SPIE.7439E..0FK,2012SPIE.8442E..1PK}. Because \gaia\ drift-scans the sky continuously, the 
CCDs operate in time-delay-integration (TDI) mode. Science requirements and engineering trade--off resulted in an optimised design with a
scan period of~6~hours, angular pixel size in the scan direction of~58.933~mas~pix$^{-1}$ , and a TDI
period of~0.9828~ms. 
The layout of devices on the \gaia\ focal plane is described in detail elsewhere (see Fig.~1 
of \citealt{crowley16}), but briefly, the devices are arranged in~seven rows, each aligned parallel to the
scanning direction (along scan), with~17 strips, each aligned perpendicular to the scanning direction (across scan).
Strips~2 and~3 are the star mappers (SM), with strips~4 to~12 consisting of the astrometric field (AF)
devices (autonomous object detection takes place between the SM and AF1 strips). Strips~13 and~14
measure low-resolution spectrophotometery in blue and red passbands (BP and RP respectively, or~XP when
referring to both), while in rows~4 to~7, strips~15 to~17 contain the~12 RVS devices.
Because of the requirement to minimise the CCD read noise and of telemetry constraints on 
\gaia\ operations, in all devices except~SM
only a restricted set of pixels around the images or spectra of objects are collected at the CCD 
readout node and telemetered. These desired pixels are read out relatively slowly to minimise the 
read noise, while unwanted pixels are flushed rapidly, resulting in a transition from 50--100~kHz in the read, 
to 10~MHz in the flush (depending on the instrument modes) in the charge transfer in the CCD serial 
register and into the readout node. In mid-2008, on--ground testing of the \gaia\ CCDs and PEMs identified 
that the electronic bias at the PEM output was not stable across this flush/read transition as a 
result of the abrupt change in operating conditions. Another lack of stability in the bias was 
identified  arising in the changing of the phases on each line of pixels during the TDI advance. 
Because the \gaia\ CCDs have four phases, the barrier phase is advanced in~four sub-pixel steps to minimise 
the blurring of the optical image as it moves along the CCD
(e.g.~\citealt{doi:10.1117/12.2227320} and references therein).
During each advance, the readout of the serial register pauses
and then recommences, so that there are~four
of these during a single CCD line (one is at the start, immediately before the prescan pixel samples). 
Again, the operating conditions are perturbed by these pauses, causing a change in the bias. 
Figure~\ref{fig:gaiaReadoutIllustration} gives a schematic illustration of the general principles of
windowed readout as implemented in all \gaia\ science devices.
The two bias perturbation effects were called bias non--uniformity, or bias NU, arising from 
flushes and glitches, respectively. Figure~\ref{fig:test_data} provides an overview of these effects as
observed during subsequent testing campaigns on--ground. They are 
different depending on the window pattern (top), but are made more evident when aligned with the line 
advances (glitches, middle) and the start of each window read (flushes, bottom).

The effect was evident in all of the focal plane CCD--PEM pairs. However, it was particularly so for those where 
the electronic gain was highest (because the gain is applied to the bias also;
this is known as the register offset and is discussed later in \secref{sec:electronicorigin}) and the
flux levels relative to the bias lowest, {\it viz.} in the RVS and, to a lesser
extent, in XP. An investigation 
was launched into the nature and origin of the effect, which differed between the CCD--PEM pairs, 
with stray capacitance concluded to be the likely culprit. In the RVS, the performance impact was found 
to be significant \citep{AllendePrieto:09}, particularly affecting the background level subtraction and 
flux calibration, and therefore the equivalent widths of spectral lines. As the
readout pattern changes only at the boundaries where the selected window
pattern changes, and hence the transitions between flushing and reading occur~\citep{DR2-DPACP-46}, 
the data are distinguished by blocks with different bias levels, 
introducing errors in the radial velocity determination. It was found that these effects were insufficiently 
corrected by data-processing approaches that sought simply to match the continuity between blocks, 
especially for faint stars where the data are noise dominated. For astrometric measurements, again the 
impact was felt principally for faint stars, where the requirements were not met, in some cases by 
more than~$30$\% \citep{deBruijn:09}.

Given the challenging accuracy requirements for astrometric and photometric measurements, and the impacts 
on RVS noted above, other options were considered to address the impact of the bias NU. While limited 
hardware and firmware mitigations were identified within the PEMs,  these were considered not to be 
sufficiently effective, and not feasible for programmatic reasons. Instead, a twin strategy of braking samples 
and of the calibration of the effect was adopted. In the case of braking samples, the flush is terminated 
earlier, and pixels ahead of the desired pixels within the window around the object are read out at the 
read, rather than flush speed. The flush/read transitions then affect principally the braking sample, 
and as there is a rapid recovery after the transition, the bias for the desired pixels within the windows 
is less perturbed. The number of samples read, and pixels flushed, is held constant in each instrument to maintain thermal 
stability. The consequence of this is that braking samples absorb window resources that could be used for astronomical measurements, and 
could therefore limit the maximum object density that can be reached. Because the object windows have 
priority, it is not always possible to assign braking samples in high-density fields in XP. A particularly limited 
number of windows is available in the RVS, in order to slow the read frequency and hence reduce the readout 
noise as much as possible, so no braking samples are applied in that instrument, and the mitigation in this 
case depends entirely on the calibration of the effect.

\begin{figure}
\centerline{\includegraphics[scale=0.5,clip,trim=0 85 0 110]{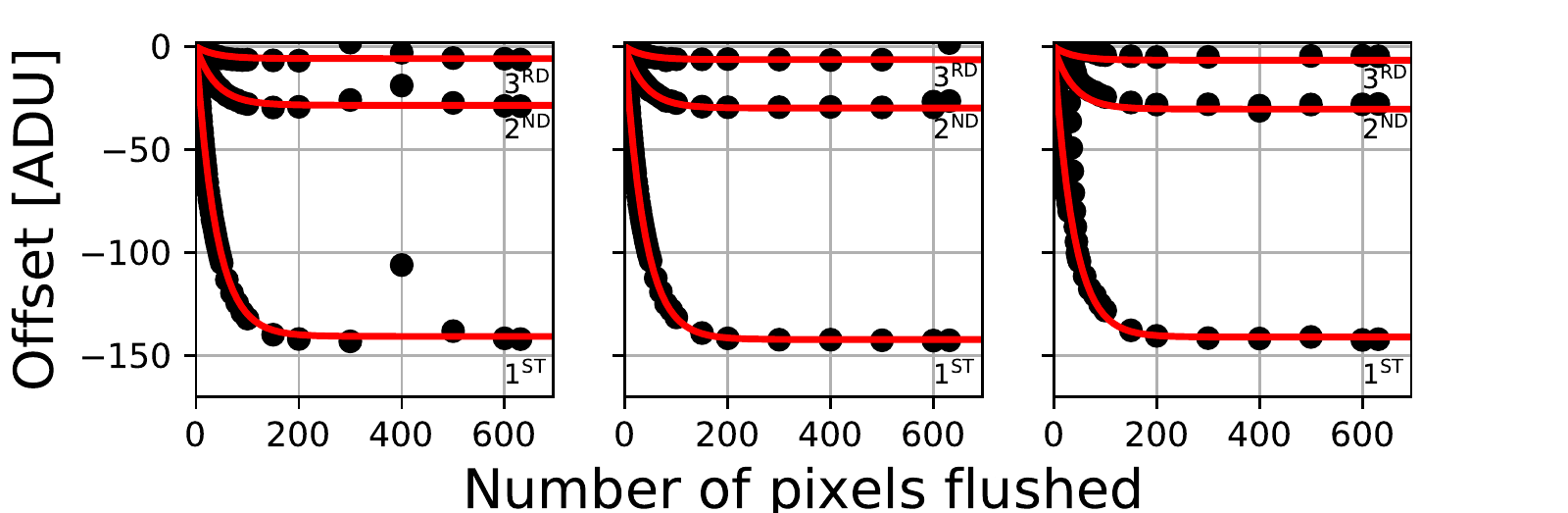}} 
\caption{Black dots in each panel show the excursion in bias level measured on--ground in an RVS test device for the first, second, and 
third pixels after the flush/read transition as a function of the number of flushes beforehand. Red lines 
show fitted analytical calibration functions for each of the three pixels. Each panel presents a different 
pattern of flushes. The saturation effect after a large number of flushes is evident, as is the rapid recovery 
in the subsequent read pixels after the first (each of the three red lines). The data shown in the panel on
the left also show the perturbation resulting from a glitch after a pause in the serial readout (see text).
Figure adapted from \cite{Boudreault:11a}.}
\label{fig:initial_calibrations}
\end{figure}

It was clear from initial investigations in early 2009 \citep{Astrium:09} that calibration was a potential 
way forward, and analytical functions were proposed to model the behaviour \citep{Astrium:10}. The depth of 
the excursion in the first read pixel after a flush/read transition was found to depend on the number of 
flushes beforehand, with the effect saturating for a large number of flushes. The time constant for the 
excursion is short, so that recovery for the subsequent read pixels is relatively rapid. 
In these measurements the effect was 
found not to depend on the flux levels in the pixels (i.e.~it was not affected by stars). The effect for 
both flushes and glitches was different for each CCD--PEM pair, and was expected to be temperature dependent. 
A larger suite of tests was carried out over~two years to characterise the bias NU behaviour, and a
task force was set up in mid-2009 (e.g.~\citealt{Hambly:09})
within the \gaia\ Data Processing and Analysis Consortium (DPAC). Analyses 
\citep{Boudreault:10, Boudreault:11a, Boudreault:11b, Boudreault:11c, Boudreault:12, Hambly:12} elaborated 
the earlier models, taking into account flushes within glitches and the behaviour of different window widths 
(encountered in the case of overlapping windows); \figref{fig:initial_calibrations} shows an example. 
These models were the basis for the codes developed within DPAC to correct for the effect 
\citep{Hambly:10}. Codes were also developed to calibrate the several dozen parameters for each model for 
each CCD--PEM pair from the use of virtual objects (these are blank windows not containing objects), and it was 
demonstrated prior to launch that satisfactory levels of correction to the test campaign data were achieved, 
with residuals at the level of $\sim1-2$ digitised units (least significant bits) even in the case of the 
RVS. Concerns remained until after launch as to the timescale of the stability of the model parameters, 
and so whether the planned monthly in--orbit calibration using special virtual object sequences 
(see later in \secref{sec:calibprocess}) was sufficiently 
frequent. However, while variations are observed in orbit, they have been found to be sufficiently gradual 
to be captured with this or slower cadence.

In this paper we discuss the on--ground treatment of the residual offset non--uniformities. Section~\ref{sec:description}
gives a description of the in--orbit measurement and on--ground calibration of the effects. Section~\ref{sec:methods} introduces the
set of models for the various components identified and gives some details of the implementation of the
calibration and mitigation software. Section~\ref{sec:results} shows results from the calibration
process in the form of calibration model residuals and also the efficacy of applying the models to mitigate
the effects in science data. Section~\ref{sec:results} also discusses the repeatability of the effects in the
context of the recalibration timescale. Section~\ref{sec:discussion} discusses the possible origin of the
effects, and we conclude this study in \secref{sec:conclusion}.

\section{Detailed description of the in--orbit electronic offset characteristics}
\label{sec:description}

\subsection{Performance measurements and requirements}
\label{sec:tdn}

\begin{figure}
\centerline{\includegraphics[scale=0.4]{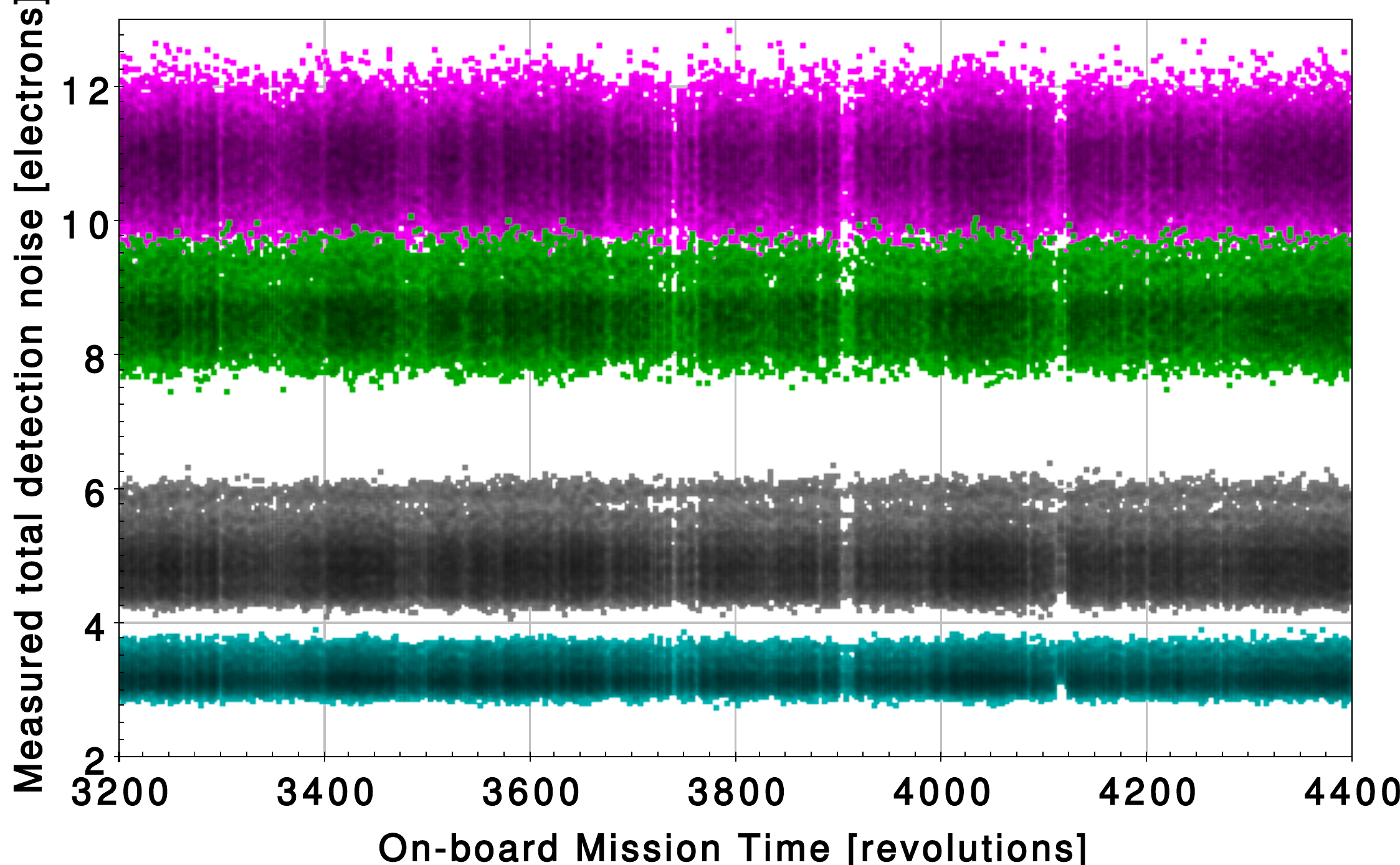}} 
\caption[]{Video-chain total detection noise measured in the initial data treatment~\citep{2016A&A...595A...3F}
from individual telemetry packets for SM (magenta), AF1 (green),
XP (grey) and RVS high resolution (cyan) versus time over an arbitrary, example~10-month period from January 2016 to October 2016 inclusive.
The horizontal axis is labelled in revolutions (units of~6~hours).
The scatter for a given instrument, i.e.~in points of a given colour, is dominated by the range in offset
level amongst the CCDs of that instrument and not instability in the offset level of any one device.
\label{fig:tdnstability}}
\end{figure}

The video--chain noise--limited performance, known as the video--chain total detection noise and incorporating effects such 
as amplifier read noise and quantisation noise from digitisation, can be measured from the sample--to--sample fluctuations on prescan
values. Each \gaia\ CCD has~14 prescan pixels available for this purpose; see for example Fig.~5 of
\cite{2016A&A...595A...1G} and~\cite{crowley16} and references therein.
Table~\ref{tab:tdn} shows summary measurements made in orbit during science observations for the various instruments and
modes in use on board \gaia,\ while \figref{fig:tdnstability} illustrates the stability of the individual 
measurements versus time for a significant portion of the same interval.
We note that the measurements for the RVS instrument are subdivided into high–resolution (HR) and low–resolution 
(LR) modes, where the latter samples are hardware binned on–chip by~3 pixels in the along–scan direction to 
reduce the read noise (and telemetry) at the expense of resolving power~\citep{DR2-DPACP-46}.
This observing mode was only used during commissioning. The high and unanticipated level of stray light 
\citep{2016A&A...595A...1G} means that all RVS observations are background noise limited (as opposed to read noise limited).  
The resulting loss of sensitivity at the RVS faint end leads to a brighter RVS limiting magnitude.  
The saving in RVS faint-end telemetry permits all RVS observations in the nominal mission to be obtained in RVS-HR, 
avoiding the loss of resolving power, dead time, and small thermal instabilities that would otherwise be introduced
into the payload as a result of the continuous sky--dependent HR/LR reconfiguration of the detectors.

\begin{table*}
\begin{center}
\begin{tabular}{lcccc}\hline\hline
\multicolumn{1}{c}{Instrument} & \multicolumn{1}{c}{Mean gain} & \multicolumn{3}{c}{Total detection noise per sample} \\
\multicolumn{1}{c}{and mode} & \multicolumn{1}{c}{ADU / e$^-$ } & \multicolumn{1}{c}{Required / e$^-$} & \multicolumn{1}{c}{Measured / e$^-$} & 
    \multicolumn{1}{c}{Measured / ADU} \\
\hline
SM & 0.2569 & 13.0 & $10.829\pm0.494$ & $2.782\pm0.127$ \\
AF1 & 0.2583 & 10.0 & $8.556\pm0.438$ & $2.210\pm0.113$ \\
AF2--9 & 0.2578 & 6.5 & $4.326\pm0.648$ & $1.115\pm0.167$ \\
BP & 0.2464 & 6.5 & $5.170\pm0.362$ & $1.274\pm0.089$ \\
RP & 0.2484 & 6.5 & $4.752\pm0.175$ & $1.180\pm0.043$ \\
RVS--HR & 1.7700 & 6.0 & $3.272\pm0.155$ & $5.791\pm0.274$\\
RVS--LR & 1.8185 & 4.0 & $2.907\pm0.177$ & $5.286\pm0.322$\\
\hline
\end{tabular}
\end{center} 
\caption[]{
CCD performance in terms of the video--chain total detection noise as measured from the second of the two prescan samples
available for each CCD, but summarised via a mean and standard deviation over all CCDs grouped by instrument.
The analysis period used was July~2014 to November~2016 and covers
the entire science operational period of DR2. Measurements are given in analogue--to--digital units (ADU) and
electrons using the gain measurements quoted in the second column, where the gain was measured on--ground 
prior to launch (this cannot be measured in--orbit owing to the non--availability of flat field illumination).
All instruments are 
statistically well within the design requirement quoted in the third column. For the RVS instrument, the measurements
are subdivided into high--resolution~(HR) and low--resolution~(LR) modes where the latter samples are hardware binned
on--chip by~3 pixels in the along--scan direction. This observing mode was used only briefly at the start of the 
mission~\citep{DR2-DPACP-46}. Note the larger gain in the RVS: this is the main contributor to the greater
impact of the bias non--uniformity in this instrument.\label{tab:tdn}}
\end{table*}

\begin{figure}
\centerline{\includegraphics[scale=0.33]{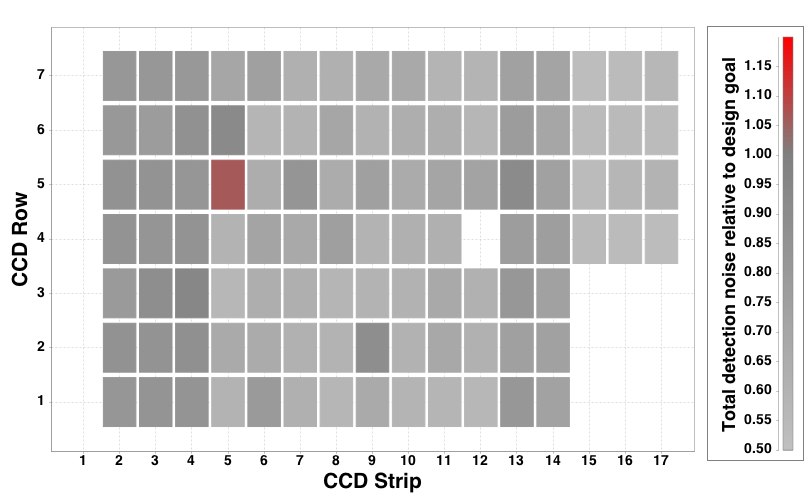}} 
\caption[]{Total detection noise measurements from sample--to--sample prescan variations mapped by colour for the
\gaia\ focal plane science devices (strip 2 = SM1; strip 3 = SM2; strip 4 = AF1; strips 5--12 = AF2--AF9; strip 13 = BP; strip 14 = RP;
strips 15--17 = RVS in high resolution mode). Only one device (AF2 in row~5) is outside the design requirement by
around 10\%; all other devices are well within their respective design requirement.
Not shown are Basic Angle Monitor and Wave Front Sensor devices in strip~1 and in strip~12 in row~4. \label{fig:tdnheatmap}}
\end{figure}

Figure~\ref{fig:tdnheatmap} displays the total detection noise measurements per device in a colour--coded heat map 
with colours mapped relative to the design requirement (column~3 in Table~\ref{tab:tdn}). 
All devices are within the requirements except for one
(AF2 on row~5), which is~$\sim10$\% outside the requirement for these devices. All others are inside their
respective requirement. RVS device video chains in particular significantly out--perform the read noise requirement
of that instrument.

\subsection{Offset characteristics during science observations}

The behaviour monitored via prescan samples as presented in the previous section is not
representative of the underlying stability in samples during
science observations. As noted above, it was determined during on-ground testing that readout 
freezes and fast flushing perturb the electronic offset at the head of each video processing chain.
The TDI time--limited requirements for the \gaia\ instruments 
makes it impossible to sample away these perturbations (with braking samples) 
and/or wait for them to recover to zero in all instruments and modes; the recovery level achieved after significant perturbation is not always the same (see below). 
Hence the approach taken (as discussed above)
is to characterise them using fixed readout patterns that cover the required parameter space and then
correct the offset excursions in software. 

Ideally, we are required to make bias measurements in darkness with zero integration time in order that thermoelectrically 
and photoelectrically produced charge are zero. Zero integration time for representative science observations (i.e.~patterns
of windows with readout sequences covering all manner of sampling and flushing relative to the fixed readout
freezes) is not possible as those observations are obtained in TDI mode. Furthermore, dark observations are not possible with 
\gaia\ in orbit since it has no shutter with which to block incident light from the focal plane array. However,
the availability of CCD TDI line gates~\citep{crowley16} allows the blocking of the transfer of thermo-- and photoelectric charge during calibration
pattern runs in TDI mode limited to a~$\approx2$~ms integration. Given the negligible dark signal~\citep{2010SPIE.7731E..1CD}, and despite the
rather high incident stray light (typically between~1 and~50~$e^-{\rm pix}^{-1}{\rm s}^{-1}$;~\citealt{2014EAS....67...69C};
\citealt{2016A&A...595A...3F}), this is a very good approximation to zero integration time under dark conditions.

\begin{figure}
\centerline{\includegraphics[scale=0.4]{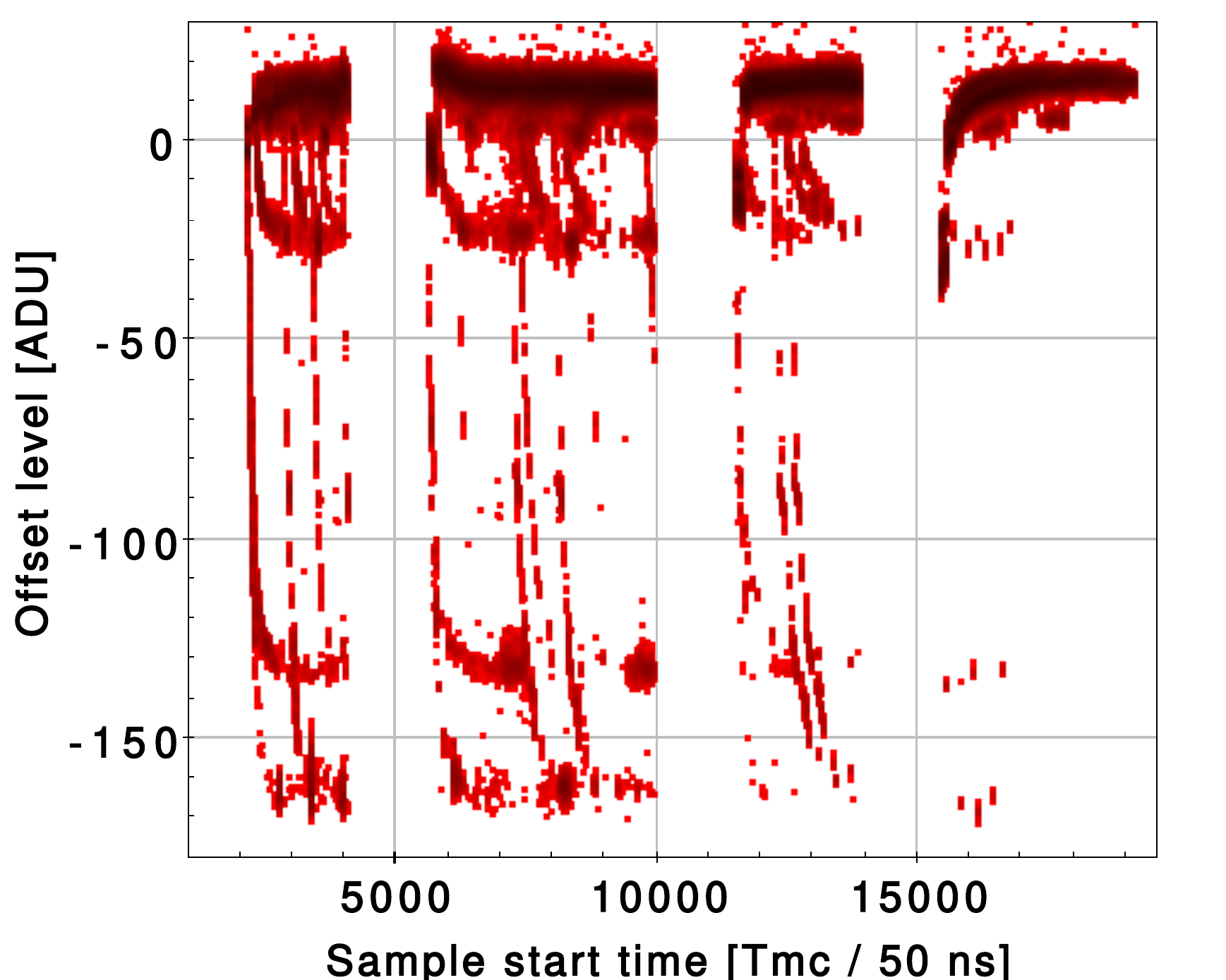}}
\caption[]{Offset relative to prescan in RVS3 (strip~17) in row~5 of the \gaia\ focal plane array as measured using 
special sequences during a calibration run in September~2016. The time axis range is just under 1~ms and
represents one TDI line in which only~72 samples can be sampled in RVS modes, but we show many samples from 
different read patterns on different TDI lines for illustrative purposes. The time axis is labelled in units
of master clock cycles (Tmc) where one clock cycle lasts 50~ns. The three empty regions correspond
to the second, third, and fourth readout pauses associated with the respective phase clock swings in the four--phase
\gaia\ CCDs described earlier in the main text. The first of these clock swings is not observed because it takes
place before the prescan samples and hence prior to image section sampling.\label{fig:rvs3row5alloffsets}}
\end{figure}

Figure~\ref{fig:rvs3row5alloffsets} illustrates the in--orbit offset behaviour relative to the prescan level in a single device. This CCD/PEM
couple, RVS3 in row~5 of the \gaia\ focal plane array, exhibits the largest offset excursions. The figure shows clear
systematic pattern noise more than an order of magnitude larger than the random sample--to--sample fluctuations seen in the
prescan samples, that is, in~the performance limit illustrated in \figref{fig:tdnstability}. 
The visible effects correspond closely to those observed before launch during on--ground testing and include
\begin{itemize}
  \item a systematic shift from zero of $\approx+10$~ADU for the greater fraction of the data;
  \item a large negative excursion after each readout freeze (the gaps at~5000,~11000, and~15000 master
  clock cycles) of several tens of ADU that rapidly recovers;
  \item a more complex systematic negative--going excursion pattern with dependency on at least one further variable that has a bifurcated, maximum offset
  somewhere between limits of $-130$~ADU and $-160$~ADU. 
\end{itemize}
The sample measurements shown in \figref{fig:rvs3row5alloffsets} are averaged over 
hundreds of repeated readouts with the same sampling and flushing pattern. This reduces the effect of individual 
sample read noise in the usual $1/\surd n$ way for~$n$ repeats.

\section{Methods}
\label{sec:methods}

\subsection{Parametric models}

The following sections provide detailed model descriptions for the calibration of the various offset 
anomaly components. We illustrate model fitting results using an in--orbit calibration run of the 
special calibration sequences in September~2016.

\subsubsection{Common baseline offset anomaly}
\label{sssec:cb}

As described in \secref{sec:tdn}, the gross electronic offset is monitored via periodic sampling of prescan pixels that precede the image section 
pixels of the serial registers in each \gaia\ CCD. The prescan samples themselves are subject to offset non--uniformity in that they
suffer the residual effects of the glitch that occurs at the the start of each TDI line scan (this is associated with the first of four
phase clock swings, i.e.~that corresponding to the missing readout pause in \figref{fig:rvs3row5alloffsets}). Hence the gross electronic offset as
measured by the prescan samples itself requires adjustment in order to correctly subtract the offset level present during the image 
line scan. This effect is termed the {\em \textup{common baseline}} offset anomaly. 

The model for the common baseline is motivated by the observation that there is in general a linear dependency on sample binning.
We do not observe, for example, any clear dependency on AC column analogous to a classical bias vector correction in the serial scan
direction in standard CCD data processing~\citep[e.g.][]{1992ASPC...23..130G}, although we
do observe low--level fixed pattern noise as a function of sample start time in the serial scan (see later). The common
baseline model is as follows:
\begin{equation}
\Delta_{\rm CB}(b_{\rm AC}) = m(b_{\rm AC}-1) + C 
\label{eqn:commonbaseline}
,\end{equation}
where $\Delta_{\rm CB}$~is the common baseline offset in ADU for a sample binned on--chip by $b_{\rm AC}$~pixels, $m$~is the gradient of the 
common baseline offset in ADU per pixel, and $C$ is the constant binning--independent offset in ADU, that is,~the 
common baseline offset for an unbinned sample.

\begin{figure}
\centerline{\includegraphics[scale=0.37,clip,trim=0 0 0 25]{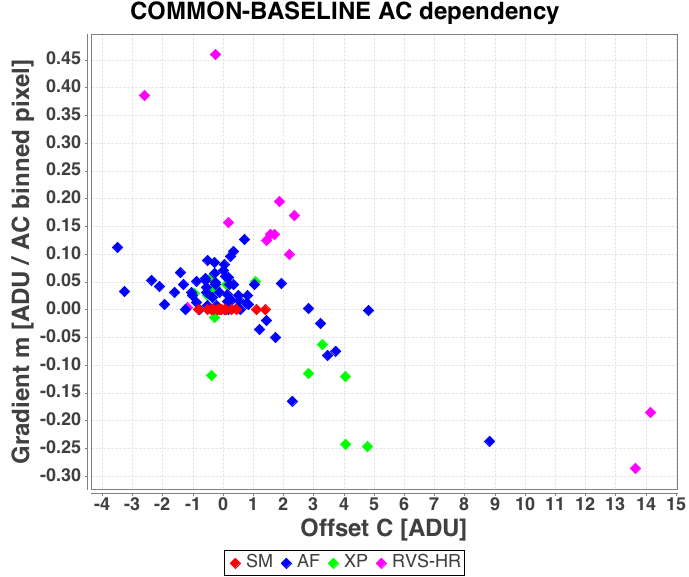}}
\caption[]{Common baseline offset and gradient ( $C$ and $m$ in \equref{eqn:commonbaseline}, respectively) for the 102 science
devices in the \gaia\ focal plane as measured in--orbit during the September 2016 calibration run. Formal error bars are smaller
than the plotted points. For all SM and AF1, there is no measurement for the gradient
as a function of sample binning because these operating modes have fixed binning of 2 pixels per sample.\label{fig:allcbs}}
\end{figure}

\begin{figure*}
\centerline{
\includegraphics[scale=0.25,clip,trim=0 25 0 0]{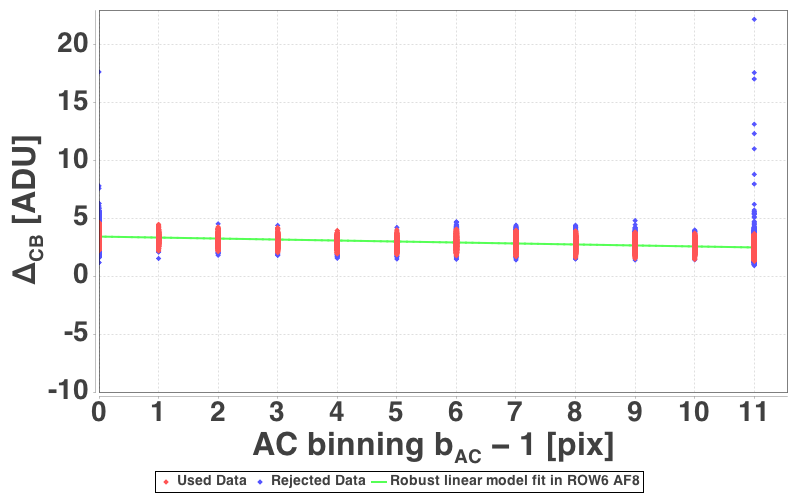}
\includegraphics[scale=0.25,clip,trim=0 25 0 0]{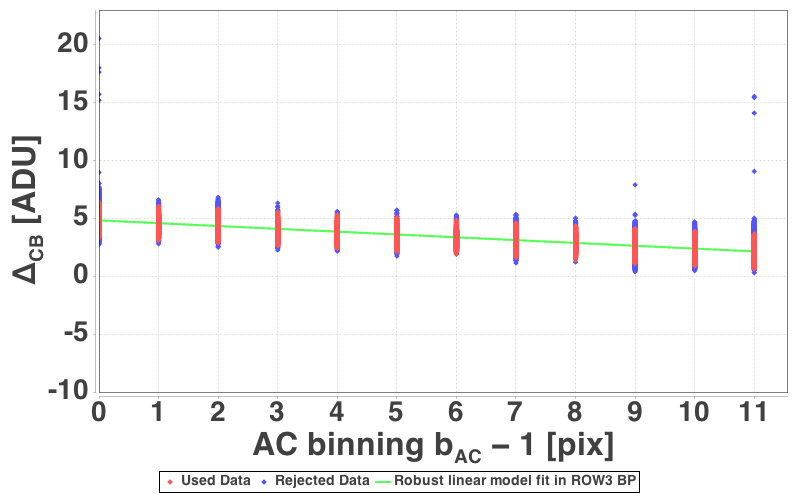}
}
\centerline{
\includegraphics[scale=0.25,clip,trim=0 25 0 0]{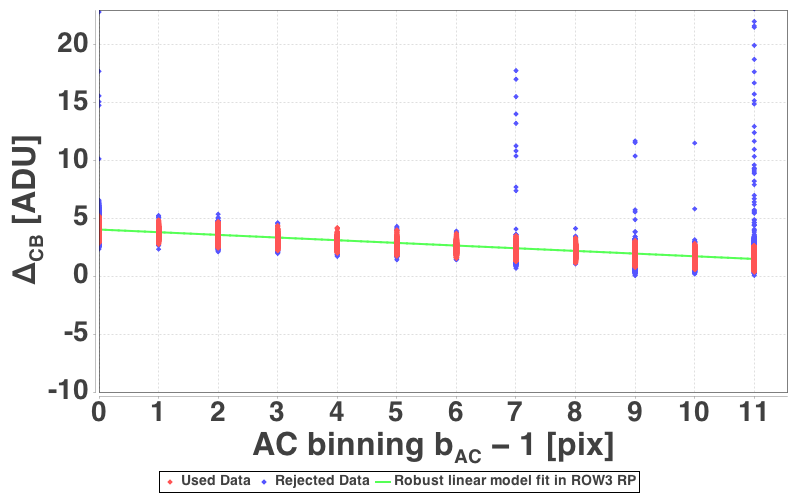}
\includegraphics[scale=0.25,clip,trim=0 25 0 0]{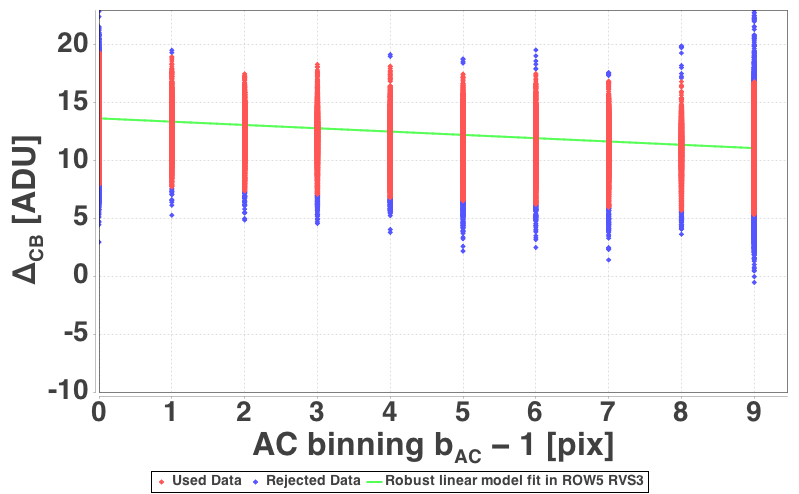}
}
\caption[]{Example common baseline fits for~4 out of the~90 devices plotted in \figref{fig:allcbs}:
AF8 in row~6 (upper left); BP (upper right), and RP in row~3 (lower left), and RVS3 in row~5, strip 17 (lower right). 
In all cases red points show calibration data, blue points
are outlying data rejected in the iterative linear least--squares fitting procedure, and the green line shows
the fitted model. Gross outliers on the positive side are the result of hot columns on the CCDs and to prompt particle (radiation
events) in their pixels, while marginal outliers on both positive and negative sides are on the
non--Gaussian tails of the sample distribution.\label{fig:egcbs}}
\end{figure*}

Figure~\ref{fig:allcbs} shows the common baseline gradient and constant for all science CCDs on the \gaia\ focal plane; 
\figref{fig:egcbs} gives some typical examples of the calibration data yielding these parameters. The
common baseline gradient is negative in many cases, providing clear evidence of a non--photoelectric, that is,~electronic, origin of the
effect (a photoelectric signal would increase with sample binning). Moreover, the highest absolute values for both
parameters are seen in RVS.

\subsubsection{Flush offset anomaly}

The flush anomaly appears as a rapidly changing (exponentially growing) perturbation to the gross offset level as a function of
the number of samples flushed immediately prior to making the science sample and stabilises at a constant
value after a characteristic timescale. The effect decays exponentially with time as samples are
read. Hence there is a sequential evolution of the perturbation as samples are read and flushed in the serial
readout. This has been modelled as
\begin{eqnarray}
\Delta_{flush}\big(N_f,b_{\rm AC}\big) & = & \big(D_{\rm BIN}(b_{\rm MAX}-b_{\rm AC})+\Delta_{flush,lim}\big)\times \nonumber \\
 & & \left[1-\exp\left(\frac{-T_fN_f}{\tau_f}\right)\right];\\
\Delta_{flush,tot}(n) & = & \Delta_{flush}\big(N_f,b_{\rm AC}\big) + \nonumber \\
 & & \Delta_{flush,tot}(n-1)\times \nonumber \\
 & & \exp\left[-\frac{T_{start}(n)-T_{start}(n-1)}{\tau_{rec}}\right].
\label{eqn:flush}
\end{eqnarray}
The fitted parameters of the model are the maximum offset variation limit for a large number of flushes, $\Delta_{flush,lim}$ 
(units of ADU); the (linear) dependency on sample binning of the maximum offset, $D_{\rm BIN}$ 
(units of ADU per binned pixel; this models the bifurcation seen in \figref{fig:rvs3row5alloffsets}, for example); 
the characteristic timescale of the onset of the flush variation, $\tau_f$ (units of master clock cycles, or~Tmc, where 1~Tmc~=~50~ns); 
and the offset recovery timescale, $\tau_{rec}$ (units of~Tmc). The independent variables of the model are $N_f$, the number of flushed 
pixels immediately before a science sample, and $b_{\rm AC}$, the number of AC pixels binned when making the sample. Fixed parameters of 
the model are $b_{\rm MAX}$, the maximum sample AC binning (12 pixels in AF and XP mode; 10 pixels in RVS mode) and $T_f$, the flush 
period (2~Tmc). In a serial sequence of samples and flushes, the total offset $\Delta_{flush,tot}(n)$ for sample~$n$ 
is the sum of the offset variation resulting from any flushes immediately preceding that sample, $\Delta_{flush}(N_f,b_{\rm AC})$, 
and the exponentially modified recovery from the total offset at sample~$n-1$, $\Delta_{flush,tot}(n-1)$. This has important
implications for the implementation of the mitigation software (see below).

Figure~\ref{fig:allflushes} shows the flush amplitude $\Delta_{flush,lim}$
and hardware binning dependency $D_{\rm BIN}$ for all science CCDs in the \gaia\ focal plane array. 
Figure~\ref{fig:egflushes} gives some typical examples of the calibration data, yielding these parameters for the devices in each
instrument strip that exhibit the highest excursion amplitudes. Once again, the RVS devices exhibit the largest
offset excursions. For AF devices, the application of braking samples
means that the full flush offset excursion is never observed, only the residual flush excursion following
recovery over the sample time of the braking sample is seen. This results in the extrapolated values of~$\Delta_{flush,lim}$ from the fits
exhibiting larger scatter, and also prevents fitting of the binning dependency parameter~$D_{\rm BIN}$ for these devices. In AF,
we fix $D_{\rm BIN}\equiv0$ for the model fits.

\begin{figure}
\centerline{\includegraphics[scale=0.37,clip,trim=0 0 0 25]{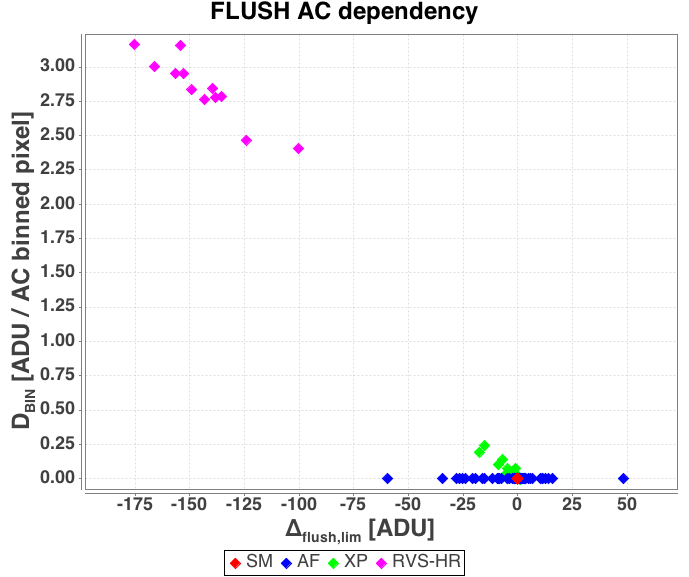}}
\caption[]{Flush-limiting amplitude and binning dependency (respectively $\Delta_{flush,lim}$ and $D_{\rm BIN}$ in 
\equref{eqn:flush}) for the science
devices in the \gaia\ focal plane as measured in--orbit during the September 2016 calibration run. 
Formal error bars are in general smaller than the plotted points.\label{fig:allflushes}}
\end{figure}

\begin{figure*}
\centerline{
\includegraphics[scale=0.25,clip,trim=0 25 0 0]{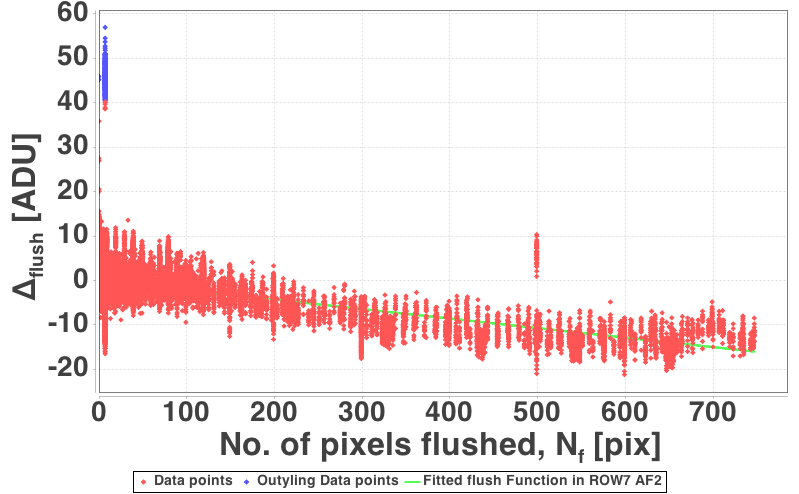}
\includegraphics[scale=0.25,clip,trim=0 25 0 0]{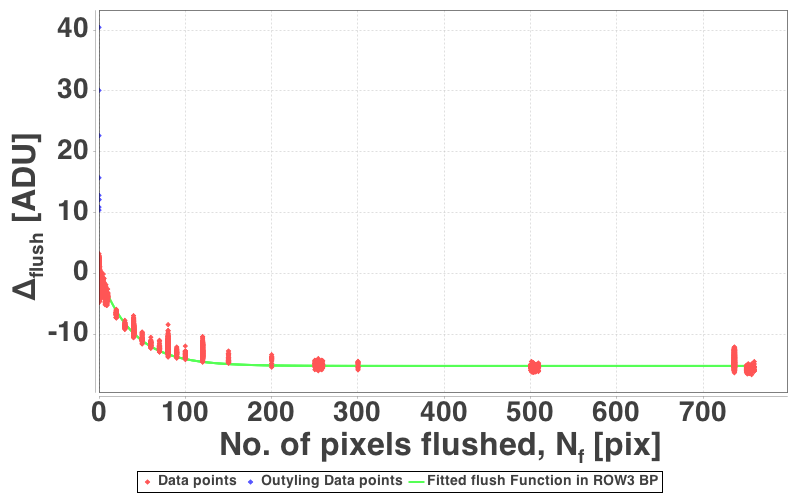}
}
\centerline{
\includegraphics[scale=0.25,clip,trim=0 25 0 0]{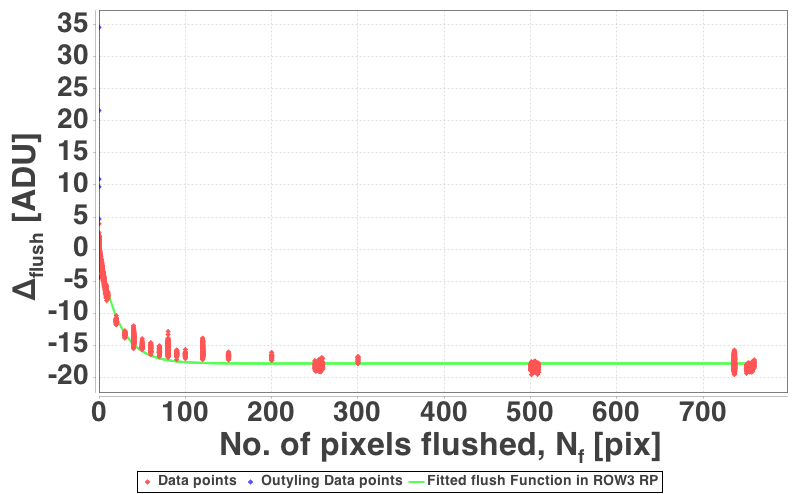}
\includegraphics[scale=0.25,clip,trim=0 25 0 0]{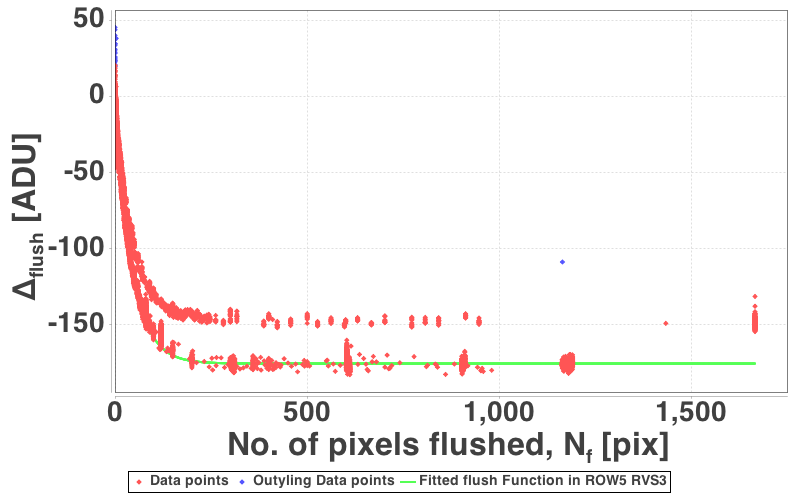}
}
\caption[]{Example flush model fits for~four of the devices plotted in \figref{fig:allflushes} with the largest
flush anomalies: AF2 in row~7 (upper left); BP (upper right) and RP (lower left) in row~3; and 
RVS3 in row~5, strip~17 (lower right). In all cases, red points show calibration data, blue points
are outlying data rejected in the iterative linear least--squares fitting procedure, and the green line shows
the fitted model (for fully binned samples; data for all binnings are shown, hence the bifurcation for high values of
$D_{\rm BIN}$).\label{fig:egflushes}}
\end{figure*}

\subsubsection{Glitch offset anomaly}

Following each readout freeze period, samples immediately following the freeze are subject to a residual
perturbation that decays rapidly as normal sample reading and flushing resumes. The first sample after the 
freeze is affected most, followed by exponentially decaying signal as further samples are read. The model
for the glitch anomaly observed in the~$n$th sample after the freeze (counting from $n=0$ for the first) is\begin{eqnarray}
\Delta_{glitch}\big(n\big) & = & \Delta_{glitch,lim} + \big(E_{\rm BIN}(b_{\rm MAX}-b_{\rm AC})+\Delta_{glitch,0}\big)\times \nonumber \\
  & & \exp \left[-\frac{T_{start}(n)-T_{start}(n=0)}{\tau_{rec}}\right] + \nonumber \\
  & & \left\{ \begin{array}{ll} 
              0 & \mbox{if $n=0$;} \\
              \Delta_{glitch,1}\exp\left[-\frac{n-1}{\kappa_{glitch}}\right] & \mbox{if $n\ge1$.} 
              \end{array}
      \right.
\label{eqn:glitch}
\end{eqnarray}
The fitted parameters of the model are the limiting value of the glitch offset far from the freeze,~$\Delta_{glitch,lim}$ 
(units of ADU); the value of the offset observed in the first sample after the freeze,~$\Delta_{glitch,0}$ (units of ADU);
the (linear) dependency on sample binning of the offset observed in the first sample after the freeze,~$E_{\rm BIN}$ 
(units of ADU per binned pixel); the value of the offset observed in the second sample after the freeze that allows
for an under--damped, or overshooting, recovery from the glitch,~$\Delta_{glitch,1}$ (units of ADU); the offset
recovery timescale,~$\tau_{rec}$ (units of~Tmc); and the characteristic overshooting recovery scale 
length,~$\kappa_{glitch}$ (dimensionless number of samples). The independent variables are the sample start time~$T_{start}(n)$
relative to the sample start time for the first ($n=0$) sample after the freeze, the sample count after the freeze ($n$), and,
as for the flush model, the sample binning~$b_{\rm AC}$. Parameter~$b_{\rm MAX}$ is fixed and has the same meaning as in the
flush model.

\begin{figure}
\centerline{\includegraphics[scale=0.37,clip,trim=0 0 0 25]{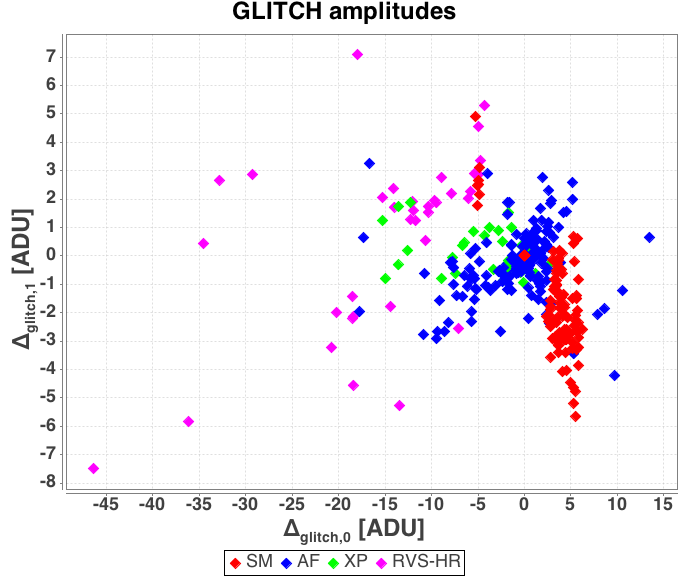}}
\caption[]{Glitch amplitudes ($\Delta_{glitch,0}$ and $\Delta_{glitch,1}$ in 
\equref{eqn:glitch}) for all glitch components in all science
devices in the \gaia\ focal plane as measured in--orbit during the September 2016 calibration run. 
Formal error bars are in general smaller than the plotted points.\label{fig:allglitches}}
\end{figure}

\begin{figure*}
\centerline{
\includegraphics[scale=0.25,clip,trim=0 25 0 0]{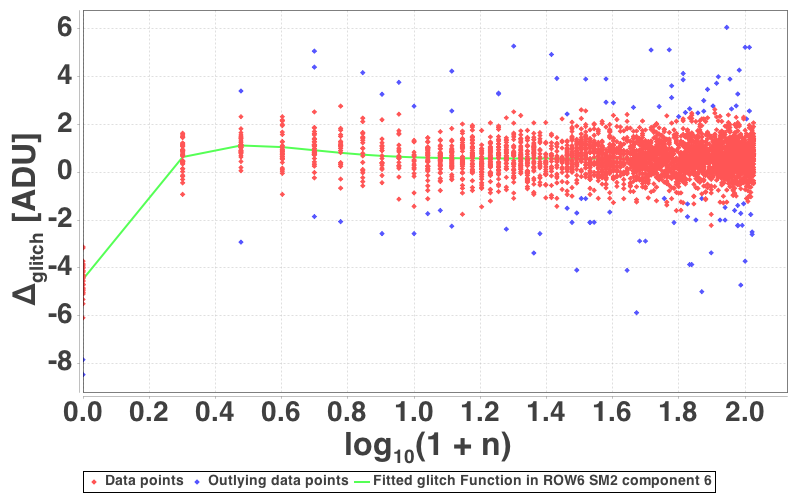}
\includegraphics[scale=0.25,clip,trim=0 25 0 0]{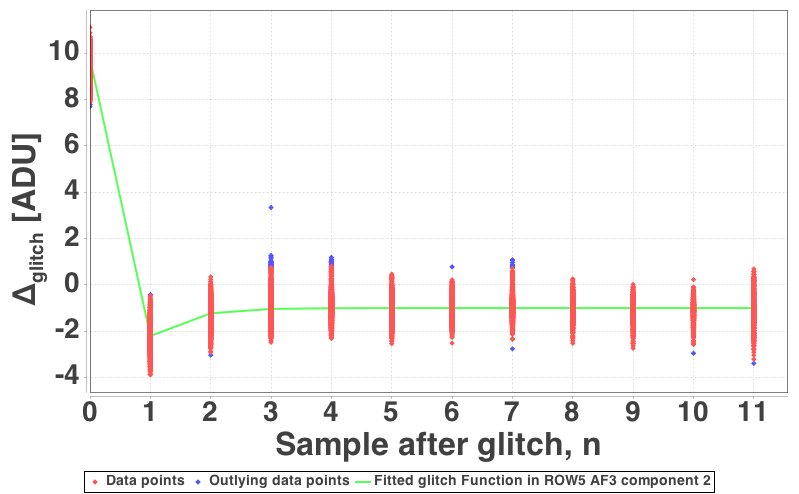}
}
\centerline{
\includegraphics[scale=0.25,clip,trim=0 25 0 0]{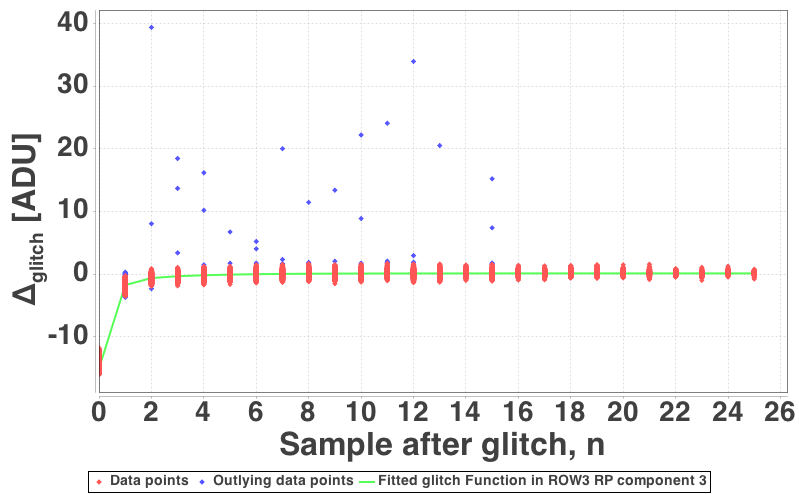}
\includegraphics[scale=0.25,clip,trim=0 25 0 0]{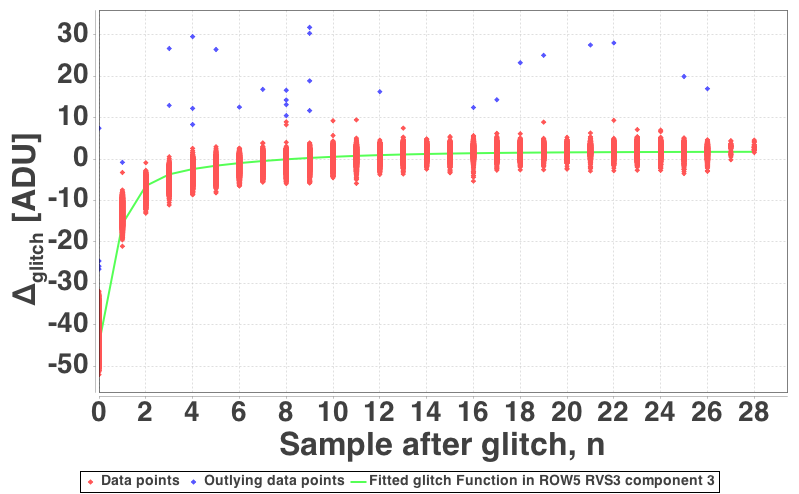}
}
\caption[]{Example glitch model fits for~four components out of all those plotted in \figref{fig:allglitches}:
glitch~6 in SM2 in row~6 (upper left); glitch~2 in AF3 in row~5 (upper right); glitch~3 in RP in row~3 (lower left); and
glitch~3 in RVS3 in row~5, strip 17 (lower right). In all cases, red points show calibration data, blue points
are outlying data rejected in the iterative linear least--squares fitting procedure, and the green line shows
the fitted model.\label{fig:egglitches}}
\end{figure*}

The glitch feature observed after each readout freeze is modelled independently for each glitch, leading to~four glitch 
components in full-resolution TDI mode and~eight in SM mode (where along--scan hardware binning by~two pixels 
enables slower readout over two TDI periods) for each CCD. The first two samples after the first glitch 
in each serial readout are always the two prescan samples that precede the TDI line scan. Hence the
prescan samples themselves are subject to an offset anomaly,
and this is the origin of the baseline offset
described in \secref{sssec:cb}. The treatment of the first glitch is then limited to fitting the
single parameter~$\Delta_{glitch,lim}$ rather than the full model.
Figure~\ref{fig:allglitches} shows the glitch amplitudes for all glitch components in all science CCDs on the \gaia\ focal plane; 
\figref{fig:egglitches} gives some typical examples of the calibration data, yielding these parameters for the devices in each
instrument strip. 

\subsection{Calibration process}
\label{sec:calibprocess}

The strategy employed to provide the data with which to determine the calibrations described above employs 
patterns of windows that are run periodically on Gaia. The windows commanded (as opposed to
being allocated based on autonomous source detection) are known colloquially as special virtual objects
in that they originate from these special calibration patterns and are devoid of any source flux.
Every~three to four~months, a set of patterns
that provides comprehensive coverage of the various offset features as a function of readout sequence timing
are run over each row of the focal plane array in turn. This involves taking each row out of science-observing mode, disabling any charge injection, disabling autonomous object detection and
confirmation between the SM and AF1
strips, and permanently activating at least gate~1 in all devices
for the duration of the run of patterns to limit TDI exposure to 2~ms. 
Pattern repeat runs and
common readout sequences contrived by spacing the windows in the along--scan direction provide multiple
sample measurements at any point in the parameter space to 
attain measurements with high signal-to-noise ratio of the offset excursions. For
an individual row, this calibration process takes approximately two hours for rows with RVS strips and around
one hour otherwise. Acquisition of a complete set of calibration data for all~seven rows of devices is
spread over a period of~ten days in order to keep commanding and downloading telemetry within the required 
operational limits. As each row is calibrated, normal science observations continue in the other rows.

\subsection{Implementation details}
\label{sec:Implementation_details}

The implementation of the calibration employs standard techniques in non--linear least--squares fitting via 
Levenberg--Marquardt optimisation (e.g.~\citealt{marquardt:1963}) of the glitch and flush parameters with
initial amplitudes of~0~ADU and e-folding scales corresponding to $100$ master clock cycles (i.e.~$5 \mu$s).
Standard linear least--squares fitting is employed for the common baseline models. All fits are made iteratively with
outlier rejection and robust estimation (median absolute deviation scaled to equivalent Gaussian sigma under
the assumption of normally distributed sample errors). The common baseline model is fitted first, followed by
the flush model and then the glitch model (XP and RVS), or glitch followed by flush (AF).
The recovery timescale parameter $\tau_{\rm rec}$ that is shared between the flush and glitch
models is determined from the flush excursions only in XP and RVS and held fixed in the glitch model fits. In 
AF $\tau_{rec}$ is determined from the glitch models only for each device and then held fixed in the flush model.
AF~and~XP calibration takes place autonomously in daily pipeline processing within the First--Look CCD one--day calibration
subsystem~\citep{2016A&A...595A...3F} with a new calibration produced only on those days where a 
focal plane row has new runs of the calibration patterns. RVS calibration takes place in the RVS daily
pipeline processing \citep{DR2-DPACP-47}. 

The application of the calibration in downstream pipelines warrants detailed description. For a given
sample in the windowed data stream, it is necessary to know the state of the offset level in
order to be able to apply the models described above. This is because there is a serial dependency 
on (i.e.~a need to compute the recovery level from) quantities such as the flush level at the
previous sample ($\Delta_{flush,tot}(n-1)$ in \equref{eqn:flush}). Furthermore, there is the need to keep track
of the number of flushed pixels between samples ($N_f$ in \equref{eqn:flush}) and the sample count after the
last readout freeze ($n$ in \equref{eqn:glitch}) in the serial scan. 

The SM CCDs are read in full--frame mode, leading to a fixed offset excursion pattern as a function of across--scan 
position in the serial scan with no flush anomalies.
Application of the calibration in SM amounts to a simple look--up of the model value from a pre-computed and stored vector
of~983 values for each of the~14 devices. 
The situation is less straightforward in the rest of the focal plane array.
Each 1~ms TDI line can have up to 24 (in AF), 71 (in XP) or~72 (in RVS) science samples with varying amounts of flushed pixels
in between as windows are allocated in the parallel scan; the sample timing with respect to the readout freezes changes
as well. Clearly, it makes sense to compute the vector of offset anomaly corrections for each TDI line in each
device once only to avoid repeating the same detailed calculations for all samples preceding the sample of interest. 
Storing these vectors and looking up values within the set would quickly become prohibitive. For example, to process
one hour of observations in~eight AF devices along one row in areas of high object density would require up to
8~CCDs$\times$3600~s$\times\approx$1000~TDI/s$\times$24~values$\times\approx$20~bytes/value (allowing for model values,
errors, and indexing overheads), which is~$\approx14$~GB.
Storing and searching the offset correction data alone in this way would be a heavy process, and this ignores all
the other processing that has to take place. In order to keep the memory footprint low while at the same time
avoiding recalculation, we take advantage of the natural time--ordering of the data stream and maintain a small buffer
of all TDI lines relevant to the window being treated at any given scan time plus one maximum window length either side.
As soon as a sample offset value is required in a given TDI line of a window, and if not already present in the buffer,
the vector of sample model offsets and errors is computed for all samples in that line and added to the buffer. The buffer
is implemented as circular first--in--first--out with fixed size, the size being chosen conservatively to be the smallest
possible while avoiding the need for any recalculation. Hence by the time a given line is overwritten, all windows
referencing that line have been processed in the time--ordered processing sweep through the data. This is a specialisation
of a generic plane sweep algorithm in data processing similar to buffered catalogue pairing (e.g.~\citealt{2005ASPC..347..346D}).
Despite this efficient implementation, a complete treatment of the offset non--uniformities is beyond the initial data treatment (IDT)
daily processing chain~\citep{2016A&A...595A...3F}, where in addition to gross prescan offset correction,
only the constant part of the common baseline, $C$ in \equref{eqn:commonbaseline}, is applied. 
RVS data are not treated in IDT but are processed in the RVS daily pipeline, where a complete treatment of 
the offset non–uniformities is performed~\citep{DR2-DPACP-47}.
For the astrometric and photometric observations, the various cyclic reprocessing pipelines apply the full 
correction~(e.g.~\citealt{2015hsa8.conf..792C,DR2-DPACP-44}).

One further complication is that the science data in themselves does not represent the totality of what was read from the
CCDs on board Gaia. This is because telemetry and on-board storage limitations sometimes result in the lowest priority
measured windows being overwritten in mass memory before it is possible to down--link these observations (this
happens mainly during Galactic plane scans). Auxiliary science
data packets are used to convey the log of all observations made on board, and these data are transmitted at high
priority. These object logs are used to reconstruct the readout history of each TDI line
of each CCD on-ground. 
One final point is that objects that are not confirmed between SM and AF1 are not recorded in these
logs~\citep{2016A&A...595A...3F}, so that no complete on--ground readout reconstruction is possible for the~seven devices in the AF1 strip. 
Offset correction in AF1 is therefore limited to gross prescan and common baseline components only, but we note 
that two braking samples are employed after each sequence of flushes, and this eliminates flush excursions.


\section{Results} 
\label{sec:results}

If left untreated, the fluctuations on sample values resulting from offset instabilities alone are large enough to degrade the video-chain performance statistics significantly. Although they are non--Gaussian, the effect of these fluctuations
can be usefully characterised by the scatter in the distribution of sample values relative to
the total detection noise (TDN) performance measured via the prescan samples (e.g.~Table~\ref{tab:tdn}). 
For the purposes of comparison against the RMS total detection noise requirements above, we now employ
an estimate of the scatter that is sensitive to any significant fraction of non--Gaussian
outliers in the sample distributions while at the same time being robust against the inevitable extremely outlying
sample values as a result of hot columns, prompt--particle events, etc. 
The estimator yields a Gaussian--equivalent RMS~$\sigma$ for a pure 
Gaussian distribution. Hence we define a robust scatter estimate (hereafter RSE) as being the difference between
the~97.5 and~2.5 percentiles (i.e.~an approximately $2\sigma$ interval) multiplied by a factor~0.2551 for a
Gaussian--equivalent~$\sigma$
(the factor comes from the quantile function of the normal distribution; cf.~the standard 
10th--90th percentile RSE defined in~\citealt{2016A&A...595A...4L}). 
For an effective
calibration and mitigation procedure, the underlying video-chain detection-noise-limited performance should be recovered
within the design requirement for the respective instrumental strips.

\subsection{Internal efficacy}

\begin{figure}
\centerline{\includegraphics[scale=0.33]{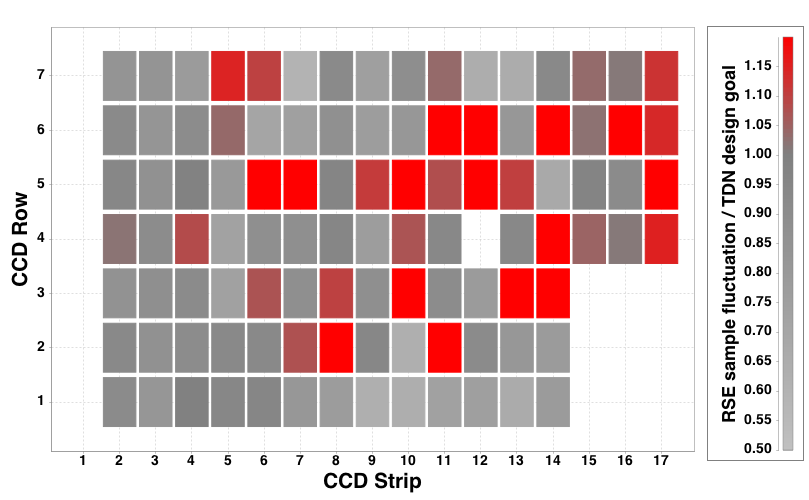}}
\caption[]{RSE (see main text) sample--to--sample fluctuations in offset instability calibrating data
colour-mapped relative to design requirements for
the \gaia\ focal plane CCDs.\label{fig:internalEfficacyBeforeHeatMap}}
\end{figure}

We first show the results of applying the calibrations to the calibrating data themselves in a necessary--but--insufficient
test of internal consistency. Figure~\ref{fig:internalEfficacyBeforeHeatMap} shows a map of the RSE
sample–to–sample variations in the raw calibrating data from September~2016 corrected for the gross 
prescan offset only and is colour--coded in the same way
as \figref{fig:tdnheatmap} to enable direct comparison. A significant fraction of devices
exhibit sample--to--sample fluctuations that manifest themselves as inflated noise with consequent performance 
outside the design requirement. Only SM and AF1 (strips~2 to~4) are relatively unaffected, partly because of a less stringent requirement
on read noise, but also because of full--frame reading (i.e.~no fast flush) in SM and the application of two sacrificial braking samples before each sequence of 
contiguous samples in~AF1.
All other strips have devices on several rows outside the design requirement. In the case of RVS, many devices are
particularly badly affected. 
Needless to say, for many devices and
under certain common circumstances (e.g.~immediately after a glitch or following a large number of flushed pixels), sample offset
excursions are much worse than the RSE performance (i.e.~several tens of times outside the TDN requirement; see for example
\figref{fig:rvs3row5alloffsets}). 

\begin{figure}
\centerline{\includegraphics[scale=0.33]{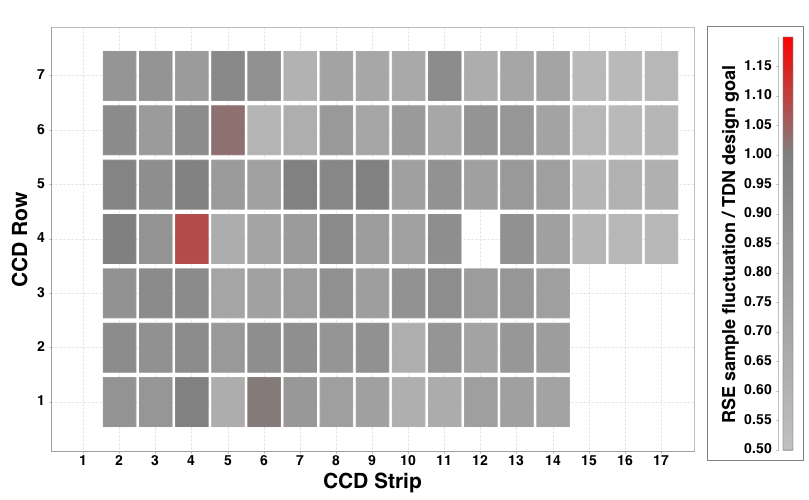}}
\caption[]{RSE sample--to--sample fluctuations in offset instability calibrating data
colour-mapped relative to design requirements for
the \gaia\ focal plane CCDs following correction by the calibrating models as calibrated from
the same data.\label{fig:internalEfficacyAfterHeatMap}}
\end{figure}

Figure~\ref{fig:internalEfficacyAfterHeatMap} shows the RSE sample--to--sample fluctuations in the calibration data after
applying the calibrated models based on them with the same colour mapping as in \figsref{fig:tdnheatmap} 
and~\ref{fig:internalEfficacyBeforeHeatMap}. The performance improvement is clear: overall, we find that 
out of the 102 science devices, 34 have an RSE performance outside the requirement before calibration and correction,
whereas afterwards, only 2 devices are outside. Even then, these 2 (AF1 in row~4 and AF2 in row~6) are 
less than 10\% over the requirement threshold.

The following sections illustrate the sample distributions before and after offset instability mitigation for the devices
that are most affected in each \gaia\ instrument. Histograms are plotted with logarithmic scaling on the $y$~-axis to
show the number and severity of non--Gaussian outliers more clearly. In each case, the upper panel shows
the sample residual distribution after prescan correction only, while the lower panel shows the same after
full correction (prescan, common baseline, glitch, and flush anomaly).

\subsubsection{RVS mode}

\begin{figure}
\centerline{\includegraphics[scale=0.2,clip=true,trim=0 25 0 25]{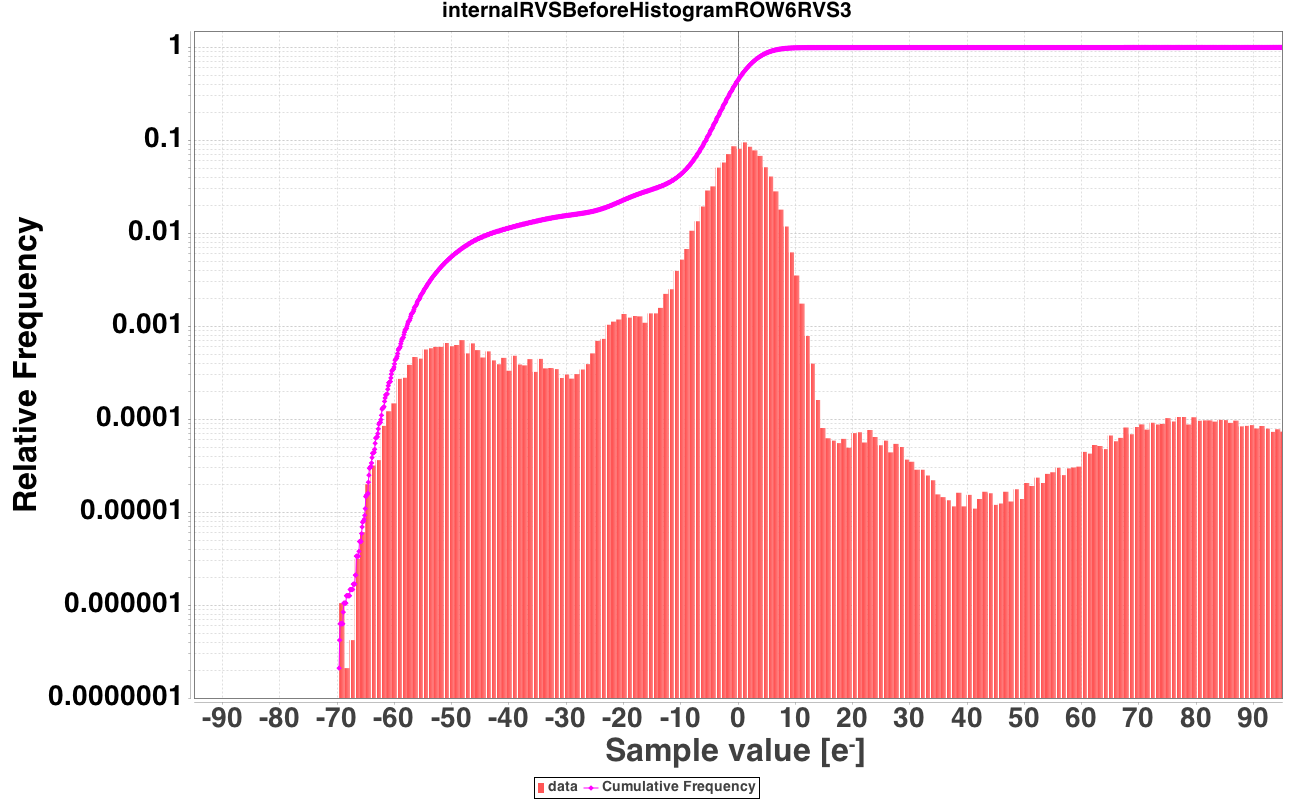}}
\centerline{\includegraphics[scale=0.2,clip=true,trim=0 25 0 25]{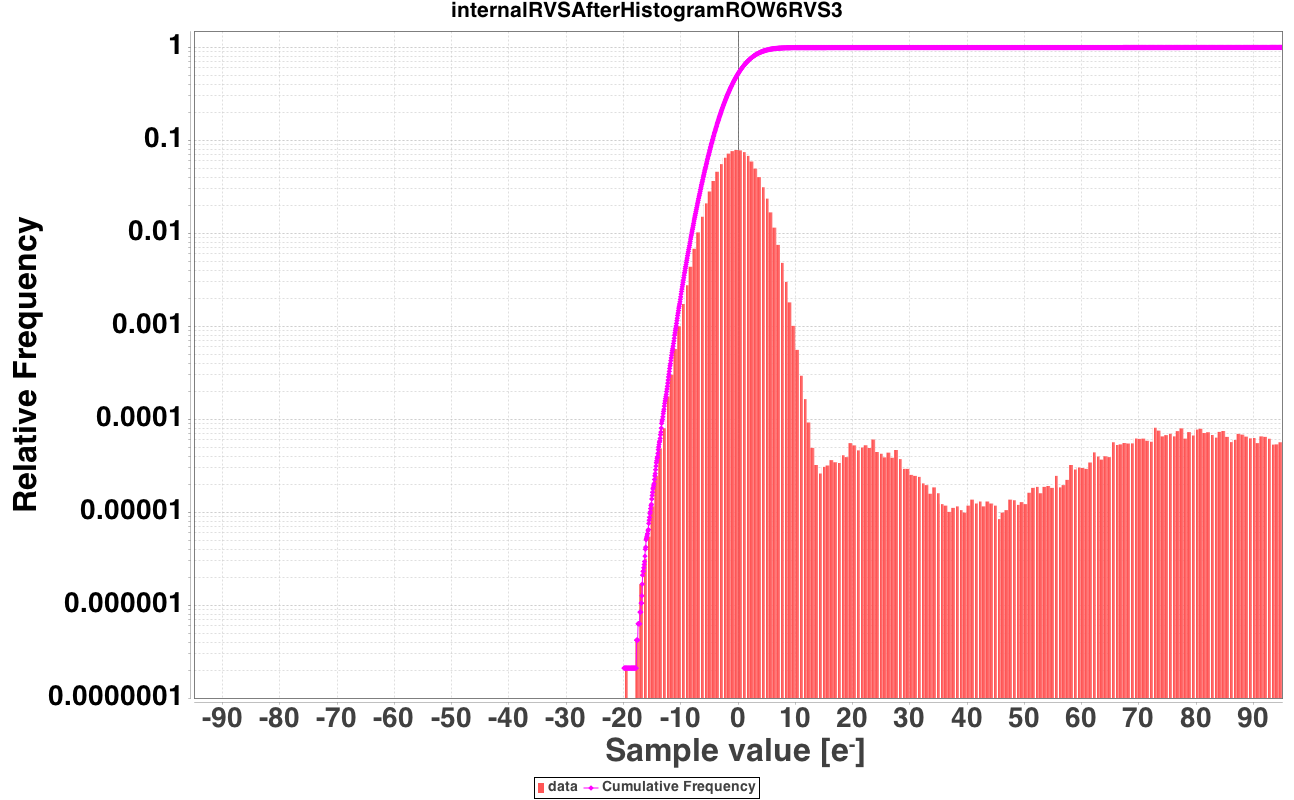}}
\caption[]{Sample distribution in the calibration data for RVS3 in row~6, strip~17 before (above) and after (below)
calibration (from the same data) and removal of the calibrated offset excursions. Red bars are histogram counts, while the
magenta line shows the cumulative count across the distribution.\label{fig:internalRVSbeforeafter}}
\end{figure}

Figure~\ref{fig:internalRVSbeforeafter} shows sample distributions for device RVS3 in row~6, strip~17. While this example is not the worst
of the RVS instrument devices, it does exhibit some prominent secondary peaks at positive signal levels.
The other~11 RVS CCDs in \figref{fig:internalEfficacyAfterHeatMap} do not have as prominent secondary peaks such that their 
before and after calibration distributions are more like those for the other instruments (see below).

The secondary peaks in RVS3 in row~6 are dominated by spurious flux (dark current), varying in strength with time, from two adjacent 
cosmetic defects in this CCD in columns 60 and 61 (see \citealt{DR2-DPACP-47} for more details on cosmetic defects in RVS CCDs).  
For the RVS CCDs, this calibration data set was obtained with all 12 TDI gates activated.  RVS3 in row~6 also has a saturating defect in 
column 199.  Figure~\ref{fig:internalRVSbeforeafter} does not include any outlying flux values from column 199.  This suggests 
that having all 12 TDI gates activated successfully prevents any gate overflow (see below).

The gate closest to the readout register is in TDI line 5.  
When this is activated, flux from lines 6-4500 will be blocked by the gate at TDI line 5, but flux from lines 1-5 will be clocked into the readout 
register.  An aluminium mask blocks TDI lines 1, 2, 5, 6, 9, and 10 so that TDI lines 3 and 4 are the only two TDI lines exposed to the sky between 
line 5 and the readout register.  For the spurious flux in columns 60 and 61 to be clocked into the readout register, defects must be in at least 
one of the TDI lines 1-5 (dark current is unaffected by the aluminium mask).
This situation is indistinguishable from every pixel in these columns being defective and producing spurious flux (hot columns).  

The spurious flux in windows including columns 60 and 61 in RVS3 on row~6 is not filtered prior to deriving the offset anomaly calibrations for 
this CCD.  This should not be necessary because the calibration fits are made iteratively with outlier rejection and robust estimation (see 
\secref{sec:Implementation_details}), although RVS3 in row~6 is the only RVS CCD to exhibit any significant coefficient variation with time.  
This is confirmed by visual inspection of the RVS3 row~6 equivalent of \figsref{fig:egcbs},~\ref{fig:egflushes}, and~\ref{fig:egglitches}.

\subsubsection{XP mode}

\begin{figure}
\centerline{\includegraphics[scale=0.2,clip=true,trim=0 25 0 0]{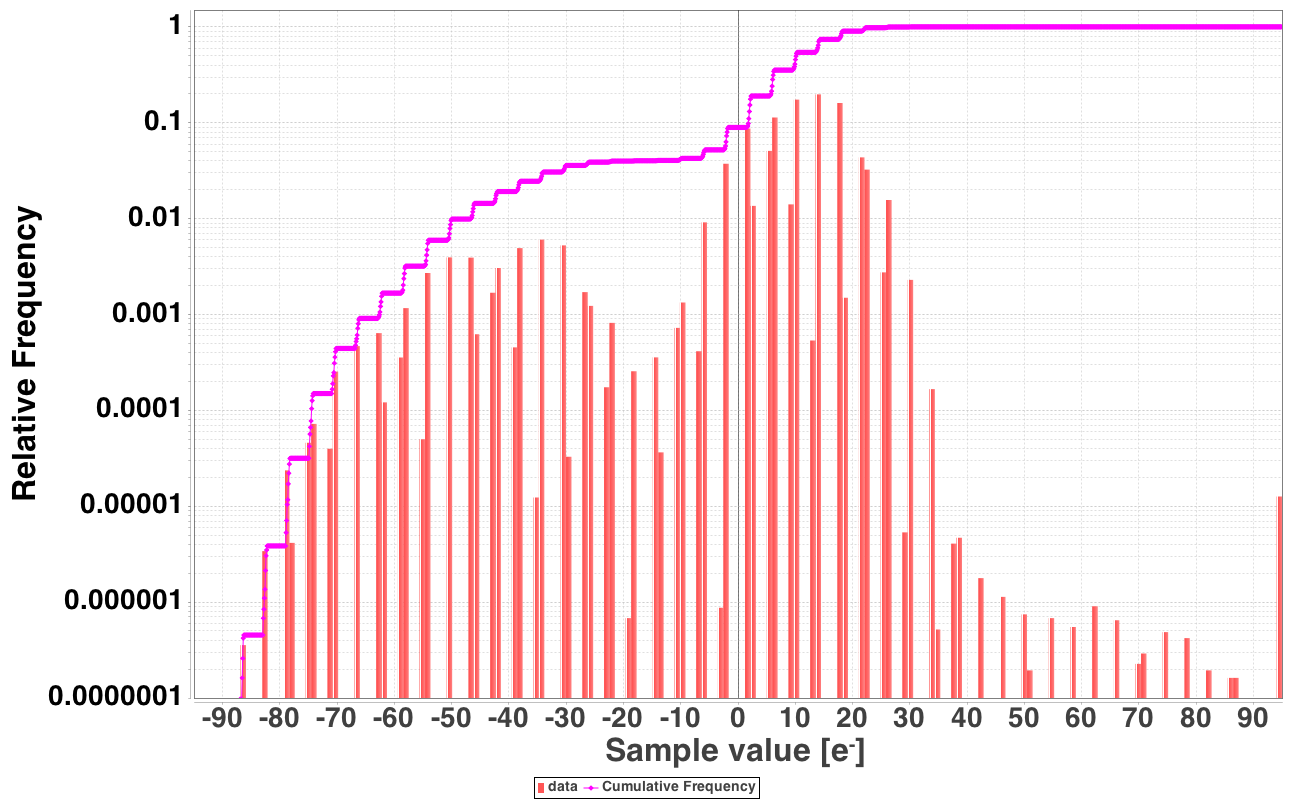}}
\centerline{\includegraphics[scale=0.2,clip=true,trim=0 25 0 0]{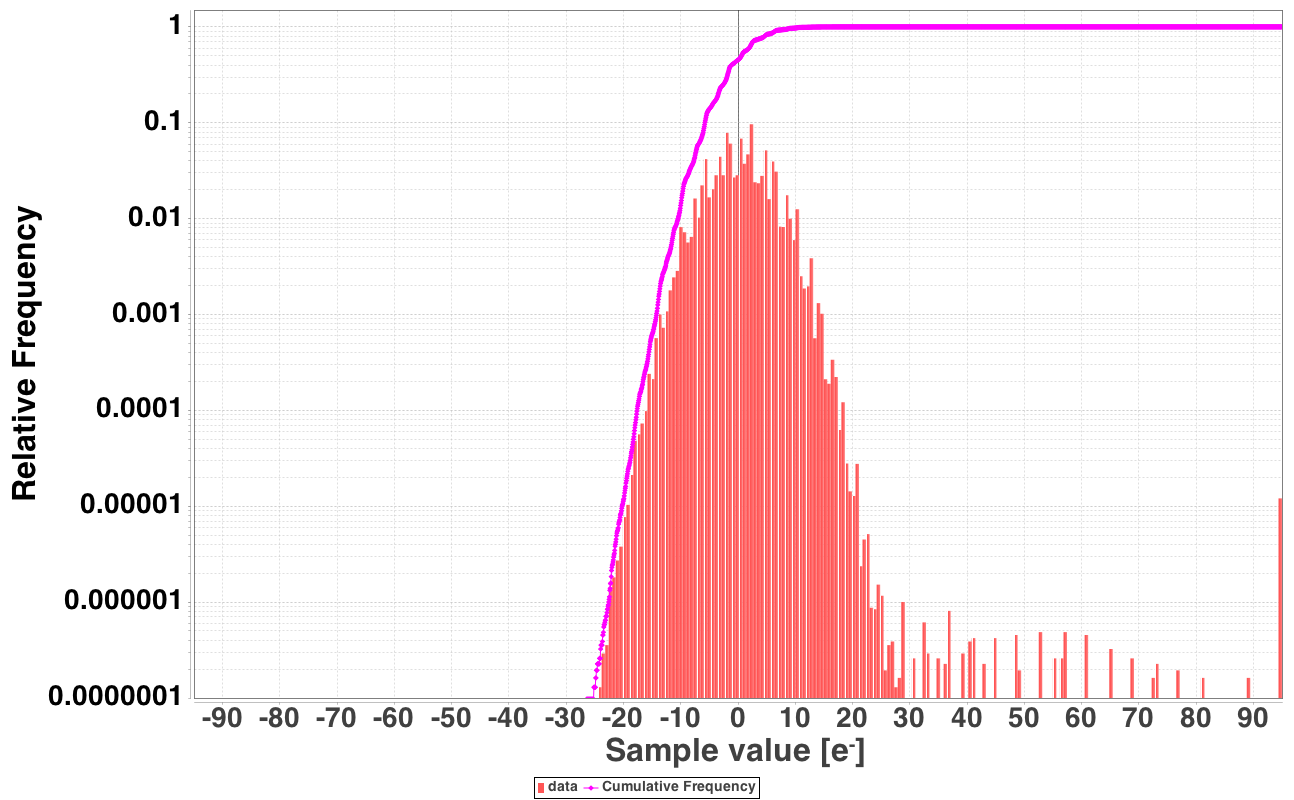}}
\caption[]{Sample distribution in the calibration data for RP in row~3 before (above) and after (below)
calibration (from the same data) and removal of the calibrated offset excursions (features and colours
otherwise the same as for \figref{fig:internalRVSbeforeafter}). Quantisation at 1~ADU (=3.9~electrons; Table~\ref{tab:tdn})
is clear in the upper panel, but slightly washed out in the lower panel as a result of applying the offset model
corrections to sample values.
\label{fig:internalRPbeforeafter}}
\end{figure}

Figure~\ref{fig:internalRPbeforeafter} shows sample distributions for device RP in row~3, which exhibits the worst offset
non--uniformity of the XP CCDs. Once again, the improvement is clearly visible.

\subsubsection{AF2--9 mode}

\begin{figure}
\centerline{\includegraphics[scale=0.2,clip=true,trim=0 25 0 0]{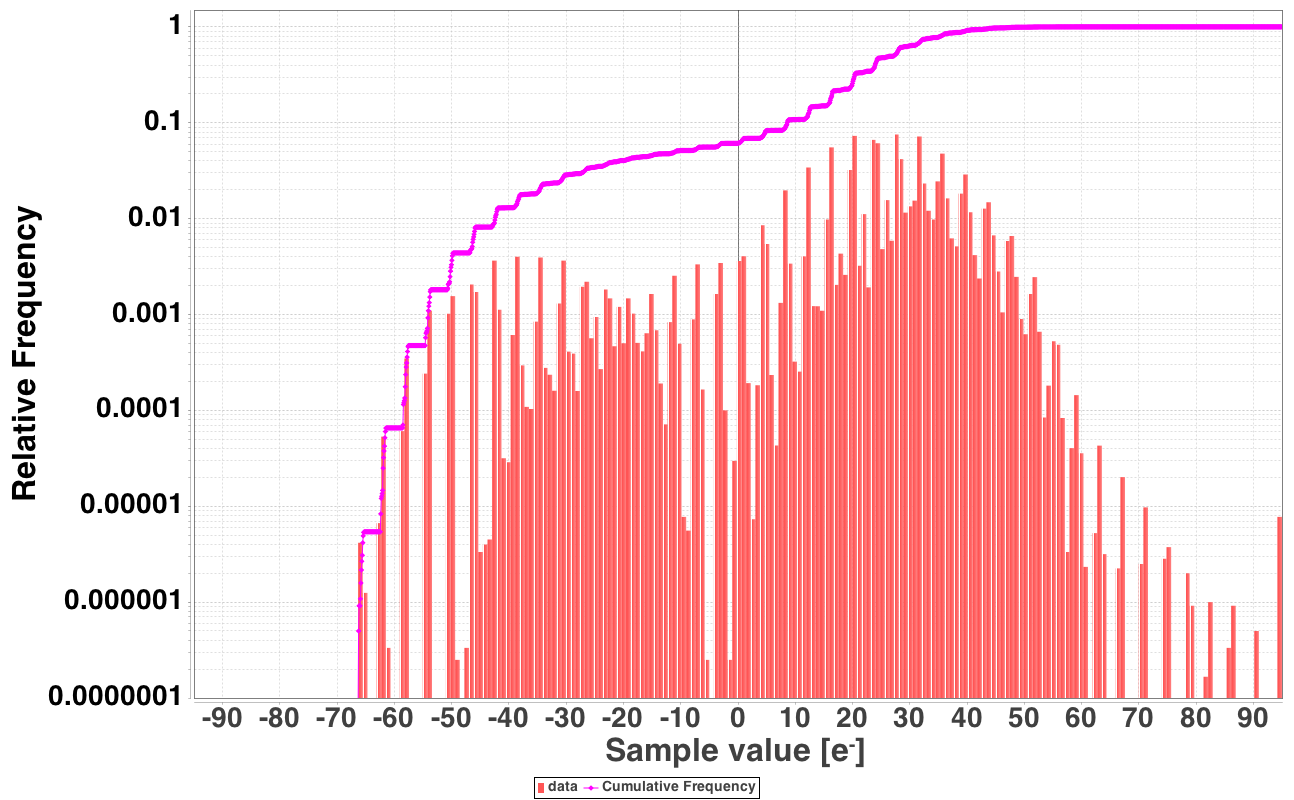}}
\centerline{\includegraphics[scale=0.2,clip=true,trim=0 25 0 0]{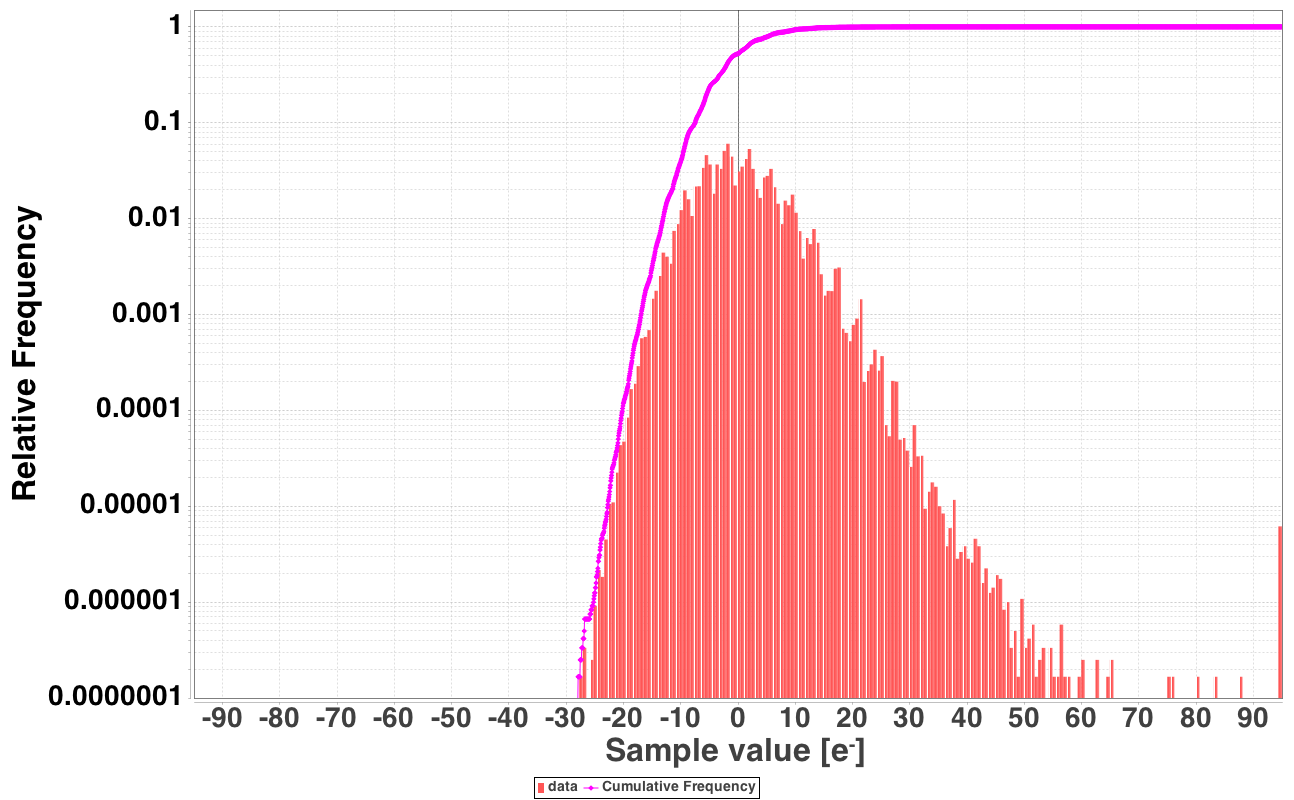}}
\caption[]{Sample distribution in the calibration data for AF4 in row~5 before (above) and after (below)
calibration (from the same data) and removal of the calibrated offset excursions (features and colours
are otherwise the same as for \figref{fig:internalRPbeforeafter}).\label{fig:internalAF4beforeafter}}
\end{figure}

Figure~\ref{fig:internalAF4beforeafter} shows sample distributions for device AF4 in row~5, which exhibits the worst offset
non--uniformity of the AF CCDs. The improvement towards noise-limited performance is clearly visible.

\subsubsection{AF1 mode}

\begin{figure}
\centerline{\includegraphics[scale=0.2,clip=true,trim=0 25 0 0]{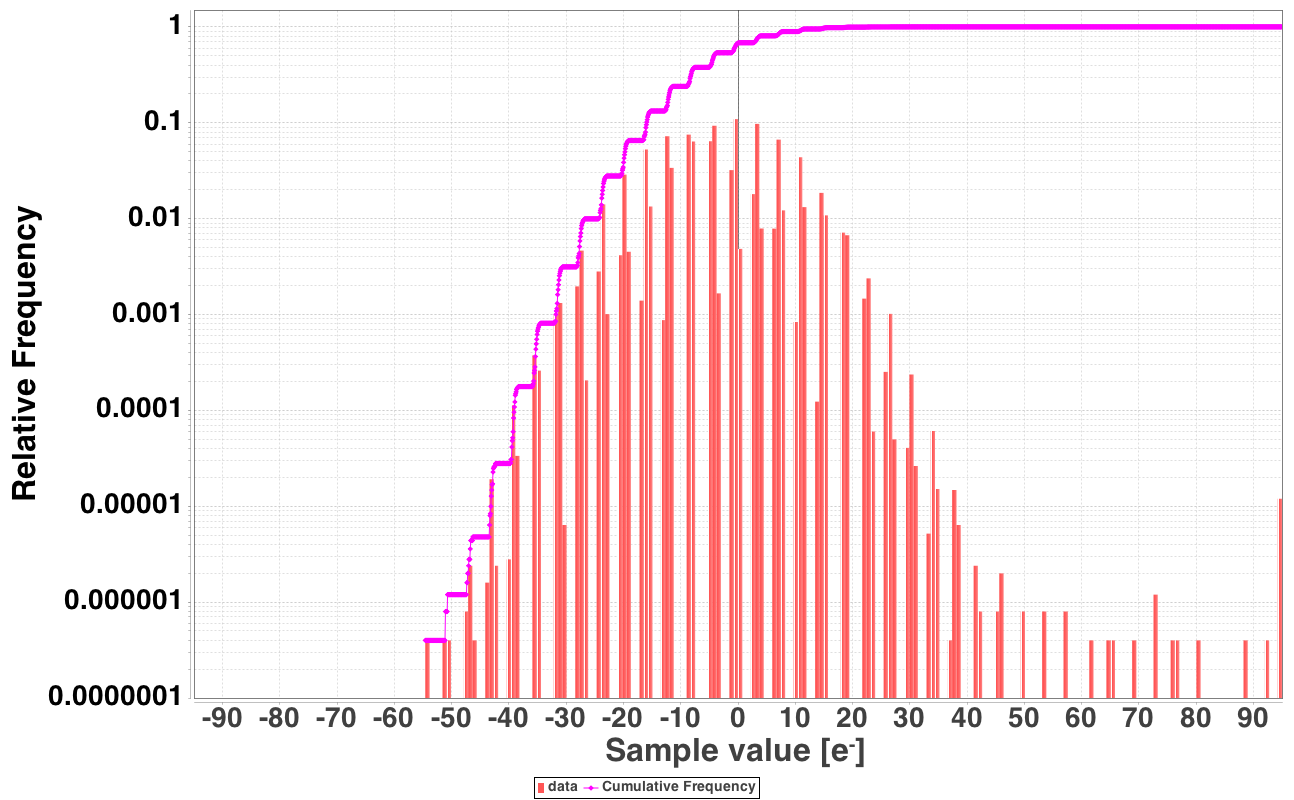}}
\centerline{\includegraphics[scale=0.2,clip=true,trim=0 25 0 0]{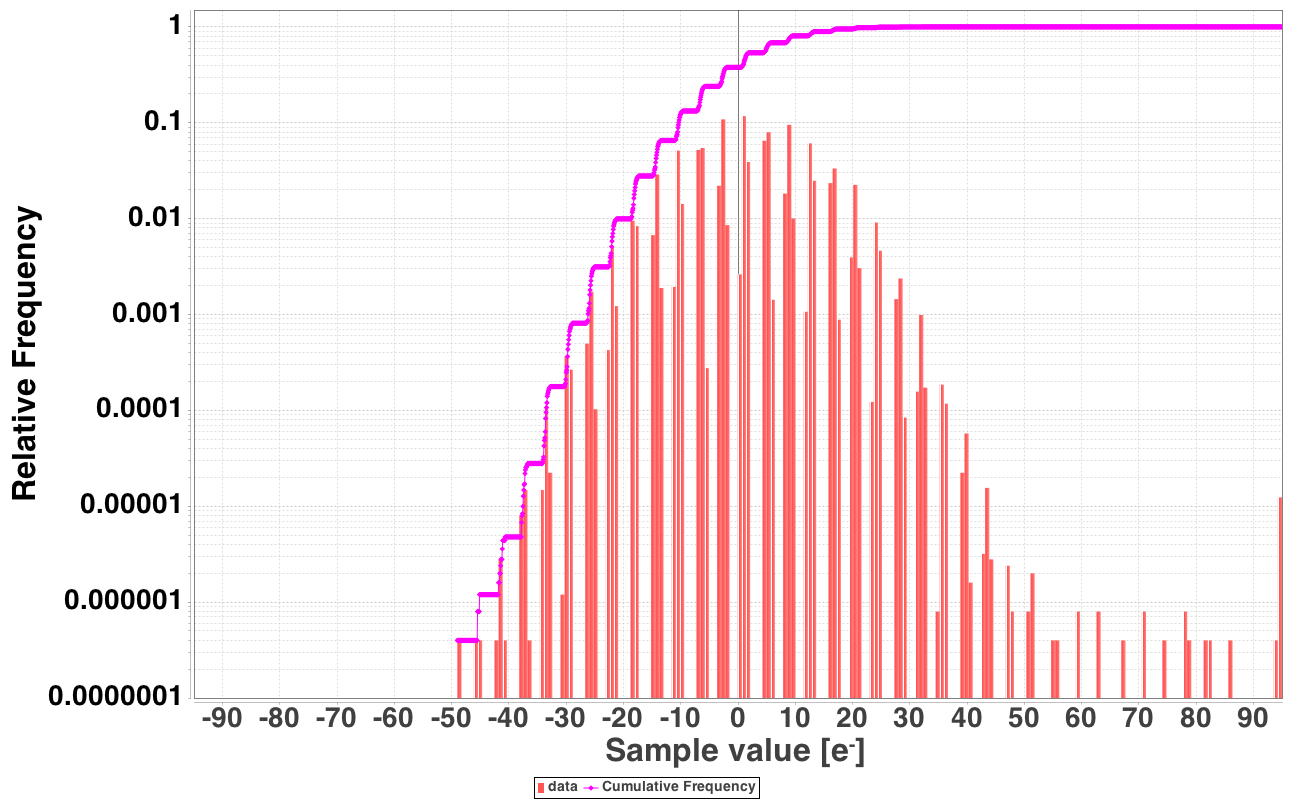}}
\caption[]{Sample distribution in the calibration data for AF1 in row~5 before (above) and after (below)
calibration (from the same data) and removal of the calibrated offset excursions (features and colours
are otherwise the same as for \figref{fig:internalRPbeforeafter}).\label{fig:internalAF1beforeafter}}
\end{figure}

Figure~\ref{fig:internalAF1beforeafter} shows sample distributions for device AF1 in row~5, which exhibits the worst offset
non--uniformity of the AF1 CCDs. In AF1, only the common baseline zeropoint correction is applied because
on--ground read--out reconstruction for science windows is not possible. 
Statistically, the improvement in calibration is marginal, being limited primarily to a better zeropoint level.

\subsubsection{SM mode}

\begin{figure}
\centerline{\includegraphics[scale=0.2,clip=true,trim=0 25 0 0]{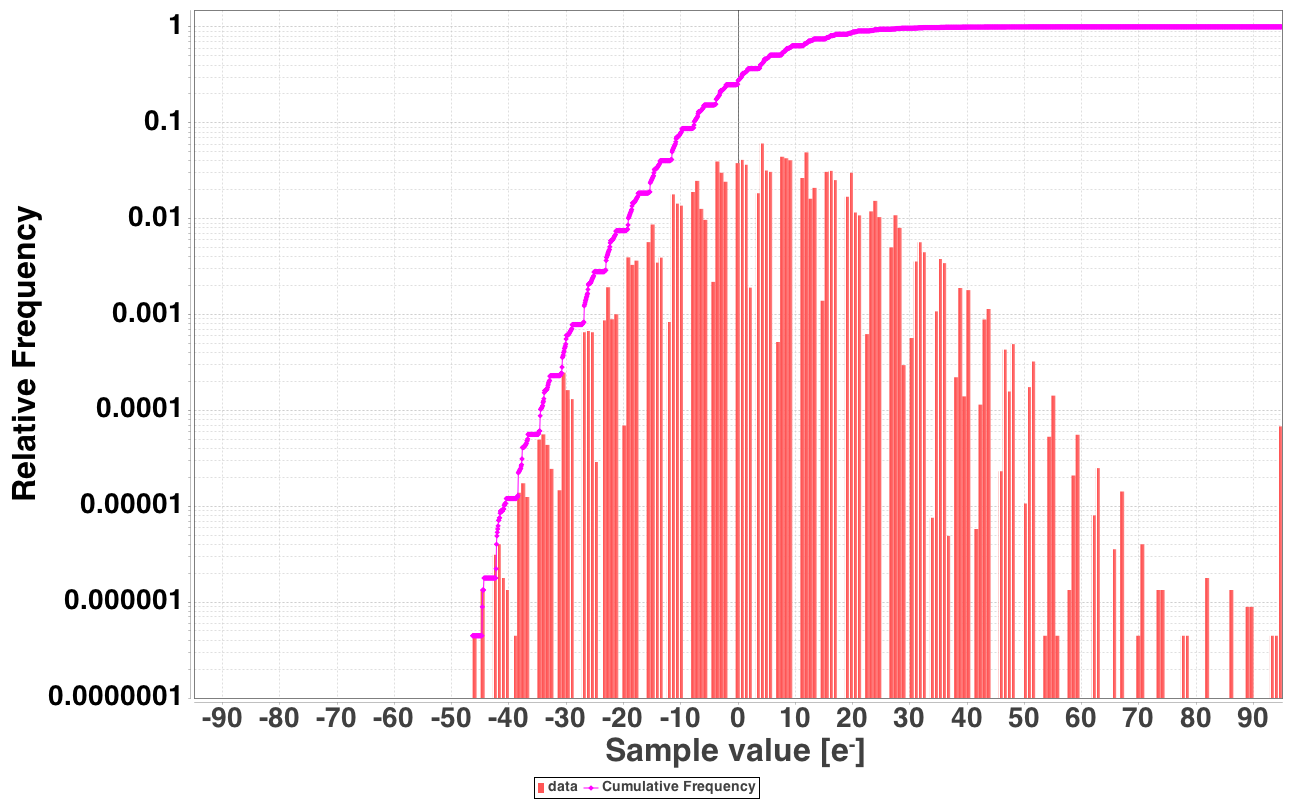}}
\centerline{\includegraphics[scale=0.2,clip=true,trim=0 25 0 0]{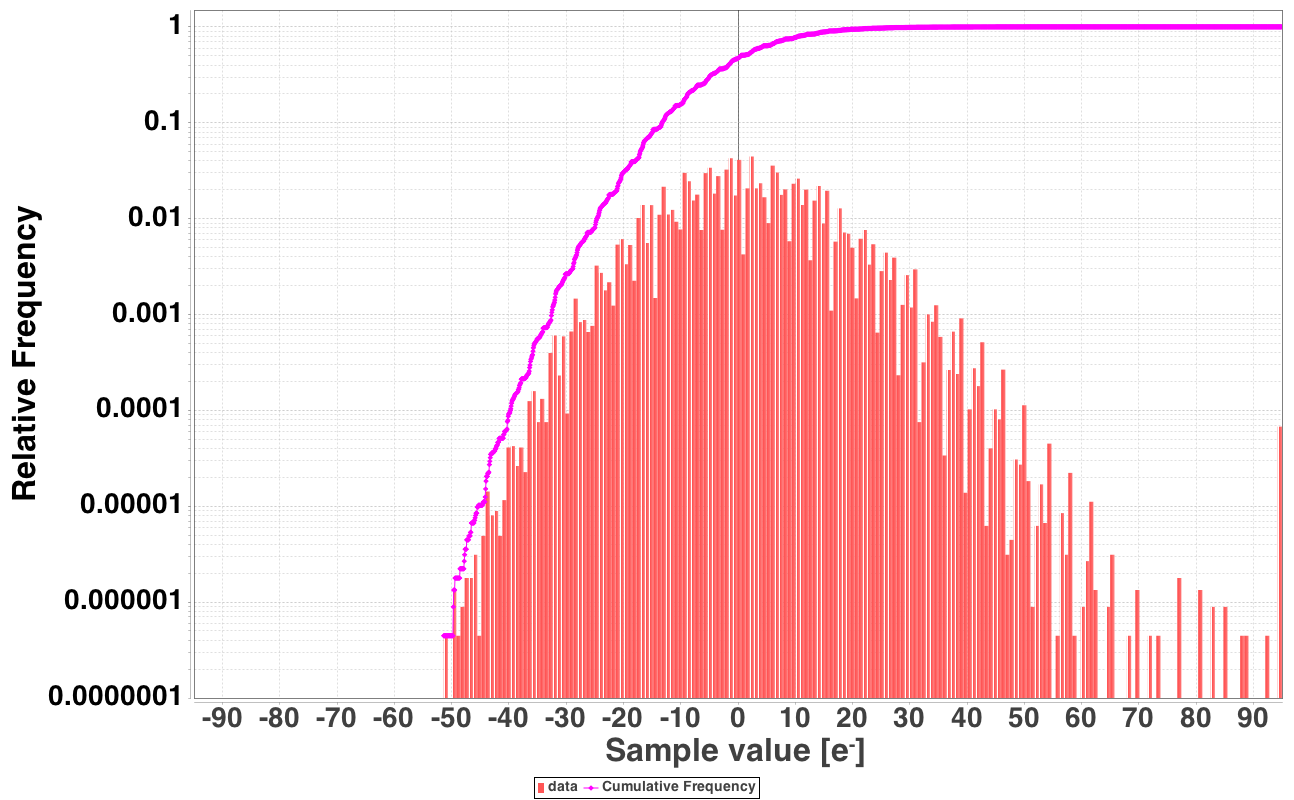}}
\caption[]{Sample distribution in the calibration data for SM1 in row~6 before (above) and after (below)
calibration (from the same data) and removal of the calibrated offset excursions
(features and colours are otherwise the same as for \figref{fig:internalRPbeforeafter}).
The data plotted exclude software-binned samples
since these samples are not used for calibration.\label{fig:internalSMbeforeafter}}
\end{figure}

Figure~\ref{fig:internalSMbeforeafter} shows sample distributions for device SM1 in row~6, which exhibits the worst 
(i.e.~highest amplitude) offset non--uniformity of the SM CCDs. 
Hence the improvement over prescan--only correction is marginal once more.

\subsection{External efficacy}

The external efficacy of the end--to--end calibration and mitigation procedure for the correction of offset
non--uniformities is best illustrated by selecting science windows from the data stream that have minimal
photoelectric signal but are otherwise affected in the same way as typical observations. To that end,
we note that the use of gates~\citep{crowley16} when observing bright stars in~AF and~XP reduces the exposure 
time of all TDI lines shared by the bright star window and all other windows that happen to be observed
very close in time on the same device but at other across--scan positions. When we select only 
empty windows (also known as virtual objects or VOs; \citealt{2016A&A...595A...3F}), 
we further limit source photoelectric flux resulting in negligible contribution to the sample fluctuations
from photon shot noise. The
shortest gate employed in AF CCDs is the fourth nearest the serial register, which has an integration time of 15.7~ms; 
the fifth nearest is the shortest employed routinely in XP, and this yields a 31.4~ms 
integration. RVS science observations are always ungated, while in SM, observations are made
with the 12th TDI gate permanently active. These configurations correspond to 4.42s full--column integrations
in~RVS and 2.85s in~SM.  

\begin{figure}
\centerline{\includegraphics[scale=0.33,clip=true,trim=0 0 0 25]{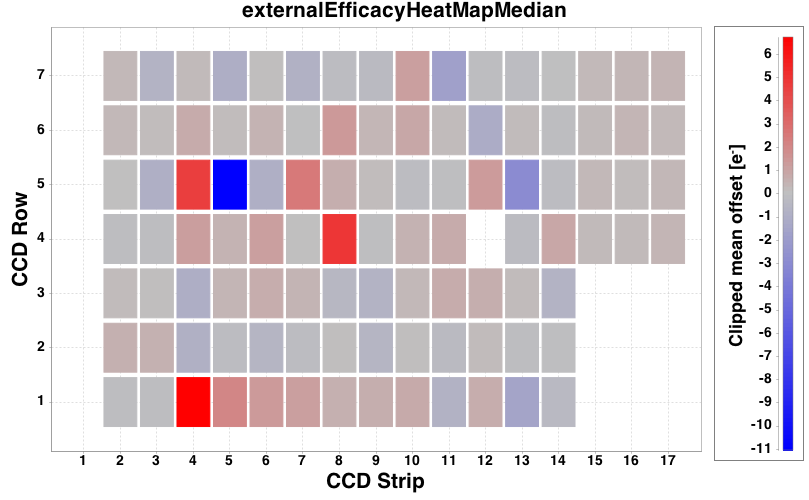}}
\caption[]{Clipped mean residuals from validation of the full offset-corrected science data employing 
empty windows, and in the case of AF and XP, short gated integrations to eliminate systematic errors
and shot noise contribution from photoelectric signal. For RVS, macrosample means were employed (see
main text), while for AF and XP, individual sample means were used.
\label{fig:externalEfficacyHeatMapMedian}}
\end{figure}

\begin{figure}
\centerline{\includegraphics[scale=0.33,clip=true,trim=0 0 0 25]{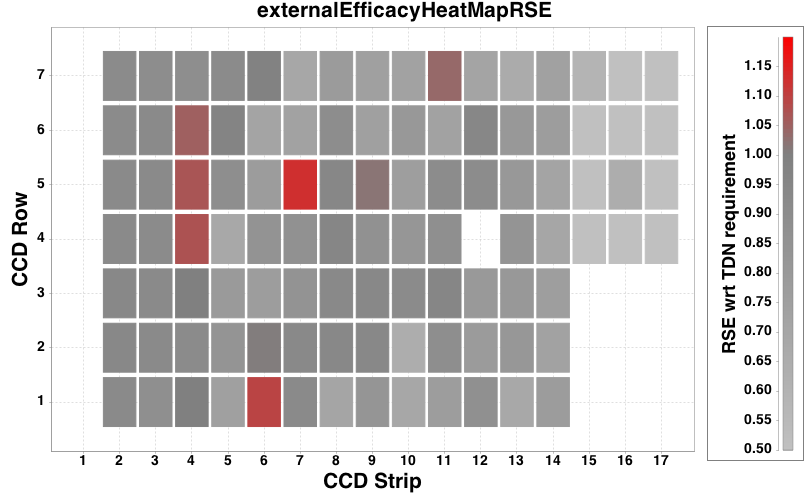}}
\caption[]{RSEs of the sample--to--sample (AF and XP) or
macrosample--to--macrosample (RVS) fluctuations in the data presented in
\figref{fig:externalEfficacyHeatMapMedian} normalised by the respective strip TDN
requirement from Table~\ref{tab:tdn}. 
\label{fig:externalEfficacyHeatMapRSE}}
\end{figure}

Figure~\ref{fig:externalEfficacyHeatMapMedian} shows the clipped mean offset of the residual sample
distribution after full bias subtraction for arbitrary sets of VOs, with integration time limited by the shortest
gate activation, in AF from OBMT revolutions~2500 to~3700 (corresponding
to observation dates from July~2015 to May~2016), and in XP from OBMT revolutions~2000 to~2300
(corresponding to observation dates from March~2015 to May~2015). The model calibration used
comes from an in--orbit calibration run during March~2015.  
The RVS data in \figref{fig:externalEfficacyHeatMapMedian} (and \figref{fig:externalEfficacyHeatMapRSE}) 
are discussed in \secref{sec:rvs}.

Figure~\ref{fig:externalEfficacyHeatMapRSE} shows the corresponding samples RSE normalised by the 
respective instrumental noise requirement for direct comparison with 
\figref{fig:internalEfficacyAfterHeatMap}. Only AF1 in rows~4 to~6 along with AF3 in row~1 and AF4 in row~5 are
marginally outside their formal performance requirement. In the case of AF1, this is likely because a complete readout history on--ground  is not available (\secref{sec:Implementation_details}).
In the case of the other two AF CCDs, we suspect that the ubiquitous application
of braking samples prevents a perfectly accurate flush calibration in all cases (of course, the benefit of the
braking samples across all of AF far outweighs this insignificant problem).
While \figref{fig:externalEfficacyHeatMapRSE}  demonstrates read-noise-limited performance
recovery for the science windows in terms of residual sample fluctuations, \figref{fig:externalEfficacyHeatMapMedian} 
indicates the presence
of an unmodelled zero-point offset (see below) at the level of~$\pm10{\rm e}^{-}$ in the devices that are affected most (e.g.~AF1 in row~1, ~AF5 in row~4, and~AF1 and~2 in row~5).

The following subsections provide examples of sample residual distributions, generally for the devices with the
largest amplitude offset excursions. The data used are 
independent of the calibrations and hence provide a true external test of the calibration and
mitigation procedure. 

\subsubsection{AF mode}

\begin{figure}
\centerline{\includegraphics[scale=0.2,clip=true,trim=0 25 0 25]{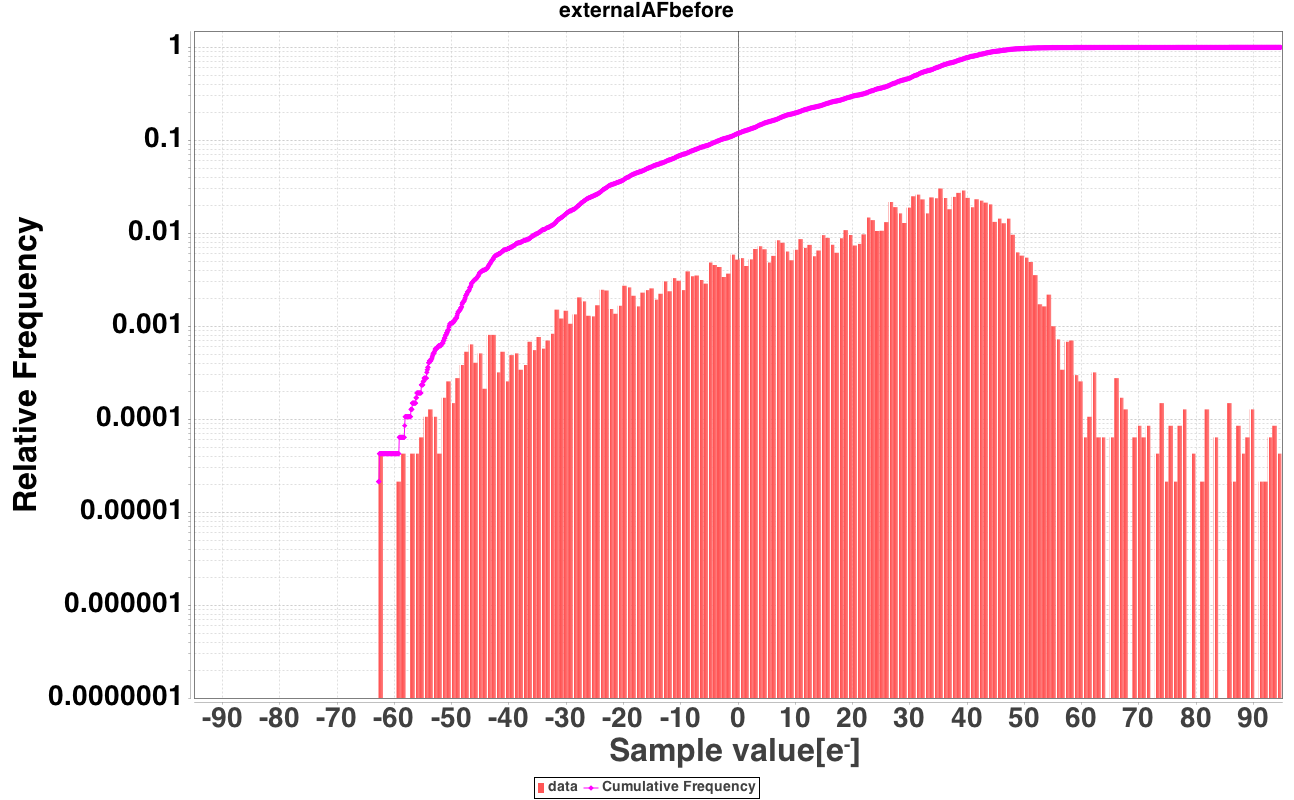}}
\centerline{\includegraphics[scale=0.2,clip=true,trim=0 25 0 25]{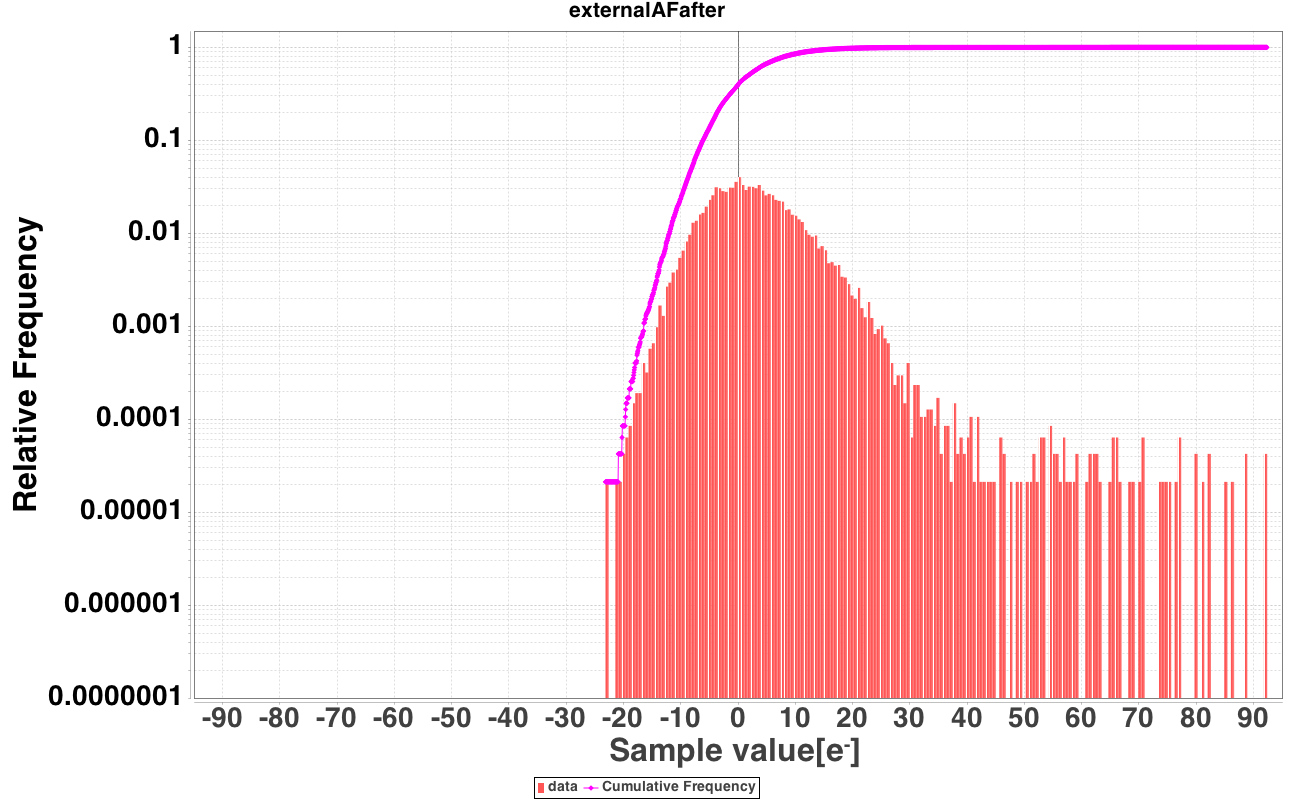}}
\caption[]{Sample distribution in the science data for AF4 in row~5 corrected by prescan level only (above) 
and with full offset corrections (below). Features and colours
are the same as for \figref{fig:internalSMbeforeafter}.\label{fig:externalAFbeforeafter}}
\end{figure}

Figure~\ref{fig:externalAFbeforeafter} shows histograms of sample values from empty windows. 

\subsubsection{XP mode}

\begin{figure}
\centerline{\includegraphics[scale=0.2,clip=true,trim=0 25 0 25]{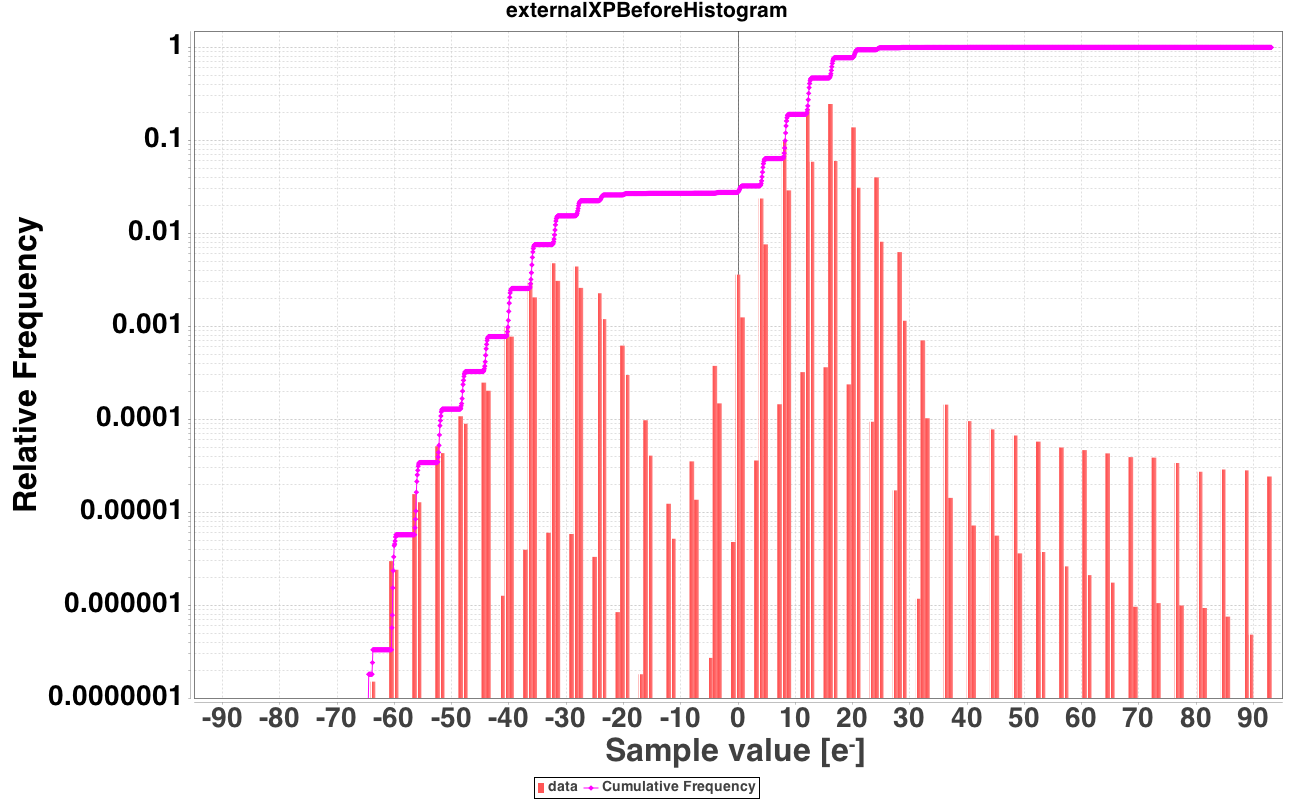}}
\centerline{\includegraphics[scale=0.2,clip=true,trim=0 25 0 25]{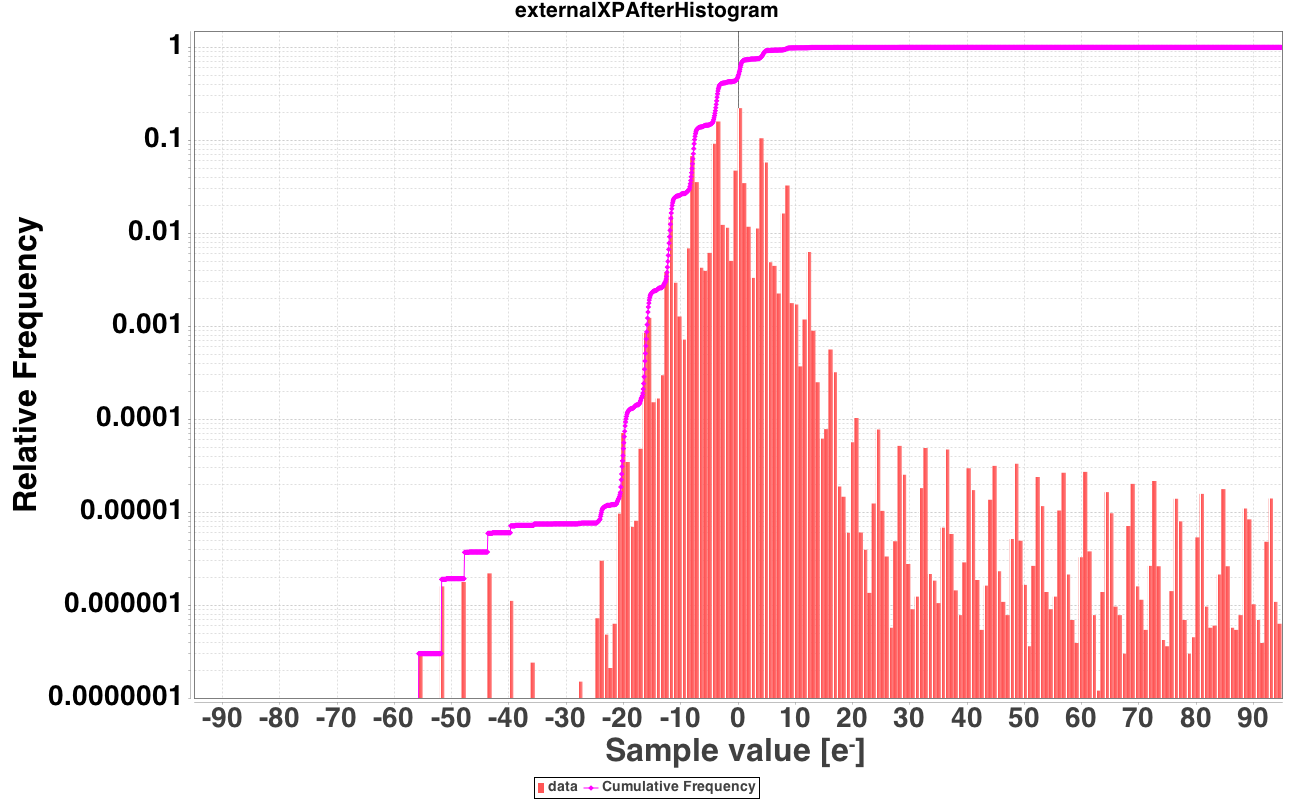}}
\caption[]{Sample distribution in the science data for RP in row~3 corrected for prescan level only (above) 
and with full offset corrections (below). Features and colours
are the same as for \figref{fig:internalSMbeforeafter}.\label{fig:externalXPbeforeafter}}
\end{figure}

Figure~\ref{fig:externalXPbeforeafter} shows histograms of sample values from empty windows for RP in row~3,
which exhibits the largest offset excursions of the devices for the~XP
instrument. In addition to low--level positive sample residuals, a very small number of negative residuals
remain, indicating imperfect bias~NU correction.

\subsubsection{RVS mode}
\label{sec:rvs}
 
The RVS VO data presented in \figref{fig:externalEfficacyHeatMapMedian} have been chosen to minimise the impact of 
stray light on the RVS external efficacy test of the bias anomaly calibrations.  They are from the 28-day ecliptic 
pole scanning law (EPSL, OBMT revolutions~1104 to~1108, corresponding
to observation dates 31 July~2014 to 1 August~2014), which have been corrected for bias prescan and non-uniformity effects 
and stray-light subtracted by the 28-day EPSL stray-light map (see \citealt{DR2-DPACP-47} for more details).  Both the derivation 
and application of the stray-light map from and to the VOs uses offset anomaly calibrations from an in--orbit calibration run 
during April~2014.  

Because RVS science observations are always ungated, all RVS VOs observed during EPSL were eligible for the external 
efficacy test.  RVS VO windows are also much longer than the AF and XP windows, which explains why the same number of RVS 
samples per device as AF and XP in \figsref{fig:externalEfficacyHeatMapMedian} and~\ref{fig:externalEfficacyHeatMapRSE} 
(30-40 million) is achieved in a much shorter time.  

\begin{figure} 
\centerline{\includegraphics[scale=0.55,clip=true,trim=0 350 0 50]{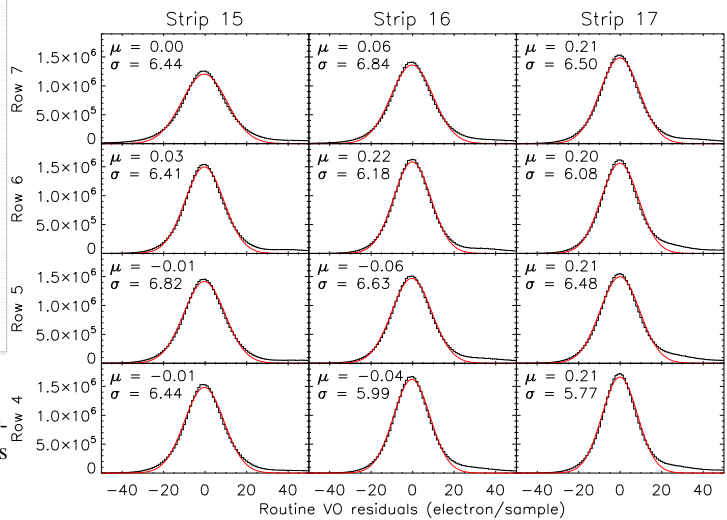}}
\caption[]{Sample distribution (shown in black) in the empty window science data (VOs) for all RVS devices 
(this data set is described in \secref{sec:rvs}). In each case, the clipped mean $\mu$ and dispersion $\sigma$ 
of the data are shown; the best-fit Gaussian distribution is overlaid (in red). \label{fig:externalRVS}}
\end{figure}

Figure~\ref{fig:internalEfficacyAfterHeatMap} shows that the RMS sample--to--sample fluctuations relative to design 
requirements are $\approx$0.5 for the majority of RVS CCDs.  This corresponds to a dispersion of $\approx$3 electrons/sample, 
which is approximately equal to the total detection noise per sample.  However, the flux--residual dispersion in the RVS 
external efficacy test data set is double this (see \figref{fig:externalRVS}).  While the stray light has been removed 
from this data set, its associated noise cannot be removed, therefore its dispersion is dominated by the stray-light Poisson noise.  
This means that sample--level RVS data cannot be used to test the external efficacy of the bias anomaly calibrations.    

In order to reduce the residual Poisson noise originating in the (high) background stray light, a clipped mean value 
within each macrosample is calculated, where a macrosample in these data contains~105 
individual consecutive TDI samples. The concept of a macrosample is described elsewhere~\citep{DR2-DPACP-46}, but for
the present purpose, we note that the readout timing of every sample within each macrosample is 
identical, resulting in the same offset excursions for each and hence the same correction model. 
Comparing macrosample means therefore reduces the background noise to a level below the video-chain
read noise but does not affect any residual macrosample--to--macrosample variations resulting from inaccurate 
treatment of the offset excursions. These residual variations are what we wish to examine for the
purpose of illustrating the effectiveness of the full offset calibration and mitigation process.  

\begin{figure}
\centerline{\includegraphics[scale=0.2,clip=true,trim=0 25 0 25]{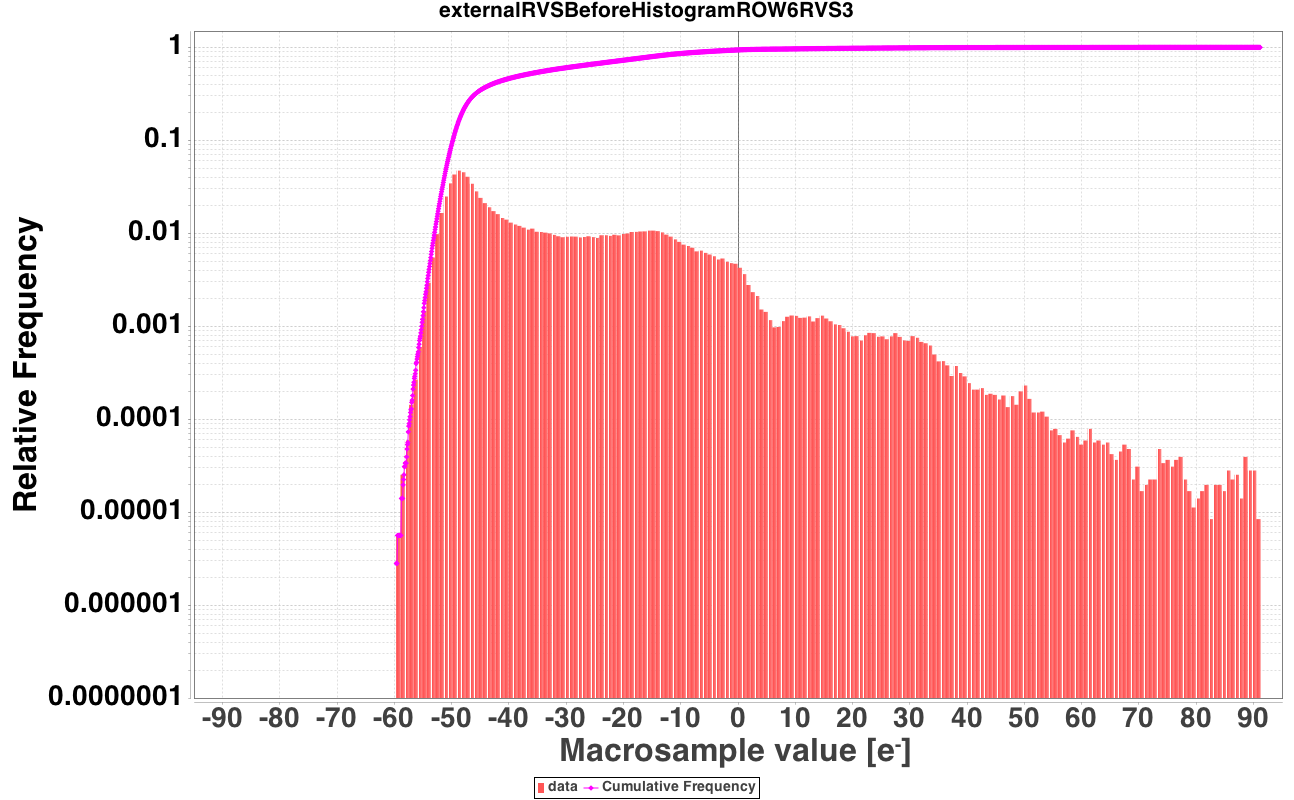}}
\centerline{\includegraphics[scale=0.2,clip=true,trim=0 25 0 25]{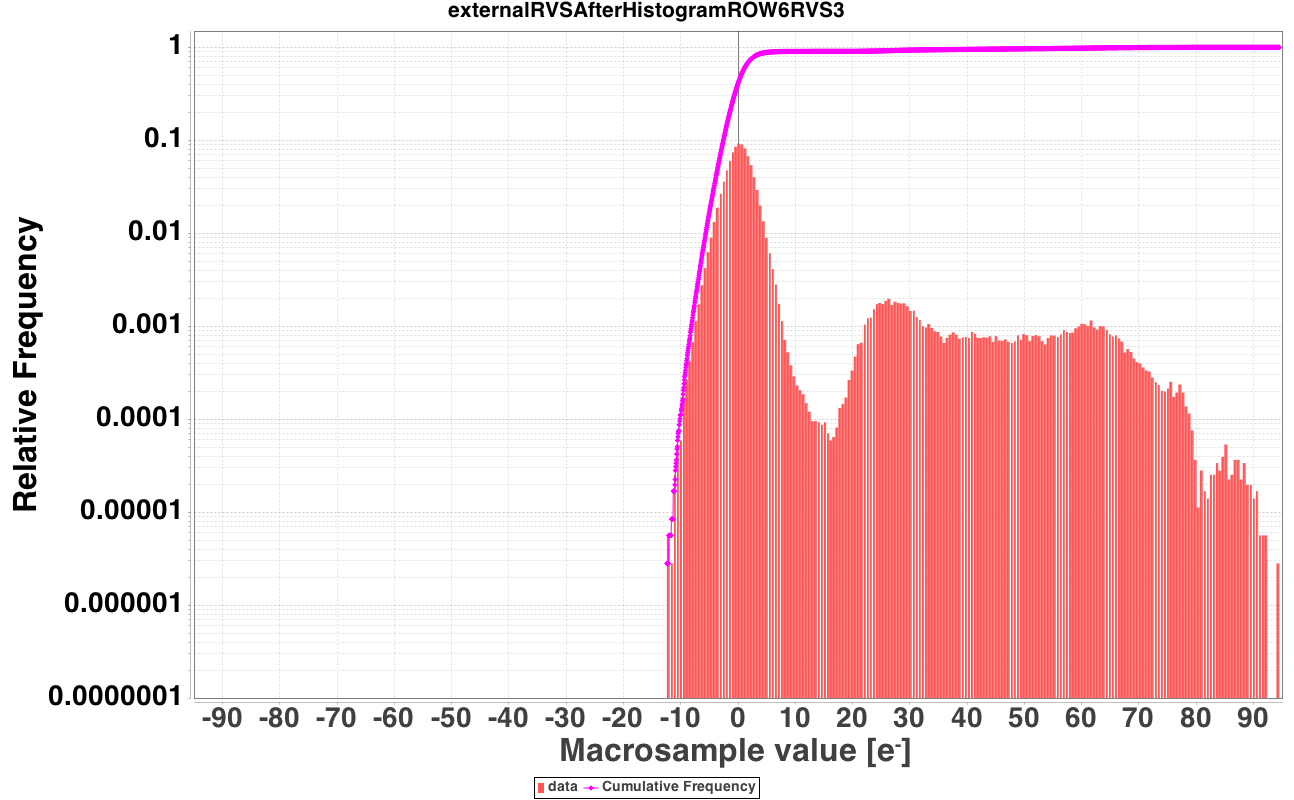}}
\caption[]{Distribution of macrosample means in science data (see text for details) for RVS3 in row~6, strip~17,
 corrected for prescan level only (above) and with full offset corrections (below).
\label{fig:externalRVSbeforeafter}}
\end{figure}

For RVS CCDs, \figref{fig:externalEfficacyHeatMapMedian} shows the clipped mean of the macrosample means for RVS VOs (1D windows).  
Each RVS CCD is close to zero, suggesting that there are no unmodelled zeropoint offsets affecting the flux in 1D RVS windows.  
For RVS CCDs, \figref{fig:externalEfficacyHeatMapRSE} shows the RSE applied to the macrosample means, where in this case
we use a standard RSE~\citep{2016A&A...595A...4L} to avoid stray-light background subtraction residuals --
RVS1 and~2 in row~7 are particularly affected by flux residuals owing to inaccurate stray-light removal (see \citealt{DR2-DPACP-46} for details).
For RVS we further limit the analysis to the negative side of the distribution in order to avoid the worst of spurious photoelectric signal.
Every RVS CCD demonstrates read-noise-limited performance recovery (relative to the total noise-detection design requirement).  

Figure~\ref{fig:externalRVSbeforeafter} shows histograms of macrosample values from VOs for RVS3 in row~6 to allow comparison with 
\figref{fig:internalRVSbeforeafter}. Figure~\ref{fig:internalRVSbeforeafter} shows that the secondary peaks have a relative 
frequency of $<$0.0001, while those in \figref{fig:externalRVSbeforeafter} are an order of 
magnitude higher.  This means that the spurious flux from defects in columns~60 and~61 is only a minor contributor to the secondary 
peaks in the latter and that they are dominated by stray-light residuals with minor contributions 
from contaminating source flux and cosmic rays.  The core in \figref{fig:externalRVSbeforeafter} is 
analysed by the RSE and is 
plotted for this CCD in \figref{fig:externalEfficacyHeatMapRSE} to illustrate the recovery of read-noise-limited performance.  
 
\subsection{Low--level systematic residuals}

The close examination of residuals following mitigation of the baseline, flush, and glitch offset non--uniformities reported
above reveals low--level systematic features in the form of residual zeropoint offsets and very low--level
non--Gaussian outliers. These appear to be fixed functions of sample start
time in the serial scan and the gate mode during calibration. The amplitude of these effects is typically
of the order of~1~ADU or lower, but in very rare circumstances,  they can be as high as~4~ADU in a few devices.
Identification, characterisation, and implementation of mitigating software in the data-processing pipelines
has come too late for treatment of these effects at \gdrtwo,\ but they are reported here for completeness
and will be removed in data released from DR3 onward.

\subsubsection{Intra--TDI phase anomaly (ITPA)}

The systematic residual component of the overall offset anomaly that is a fixed function of sample start time in the serial
scan is analogous to the low--level pattern noise often observed in zero-exposure bias frames in conventional 
full--frame CCD imaging. The origin of this component is unclear, but it is likely that the close proximity of more than 100
CCDs and their PEMs makes for an electronic environment susceptible to minor cross--talk and electromagnetic
pick--up. 

\begin{figure}
\centerline{\includegraphics[scale=0.5]{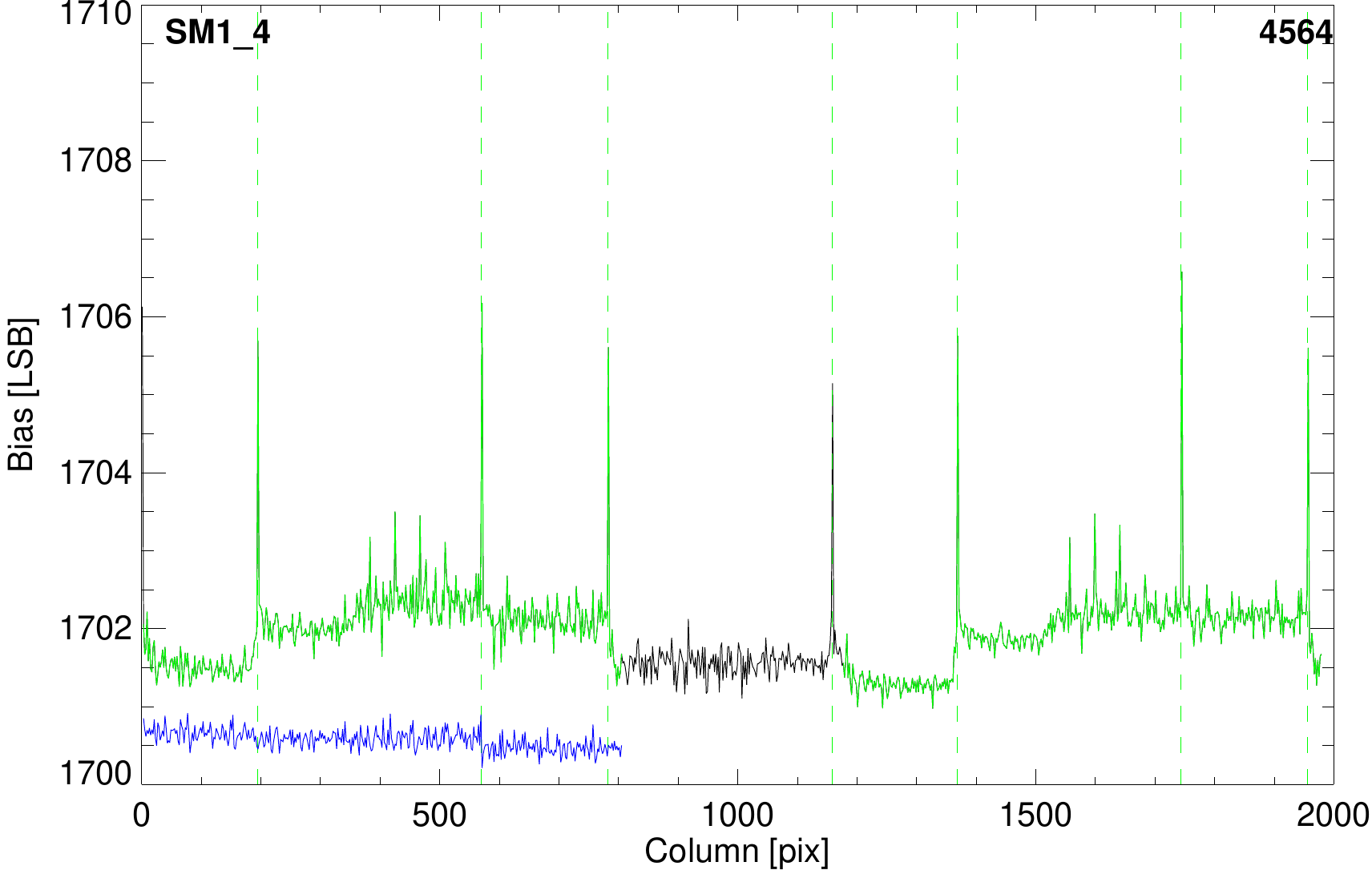}}
\centerline{\includegraphics[scale=0.5]{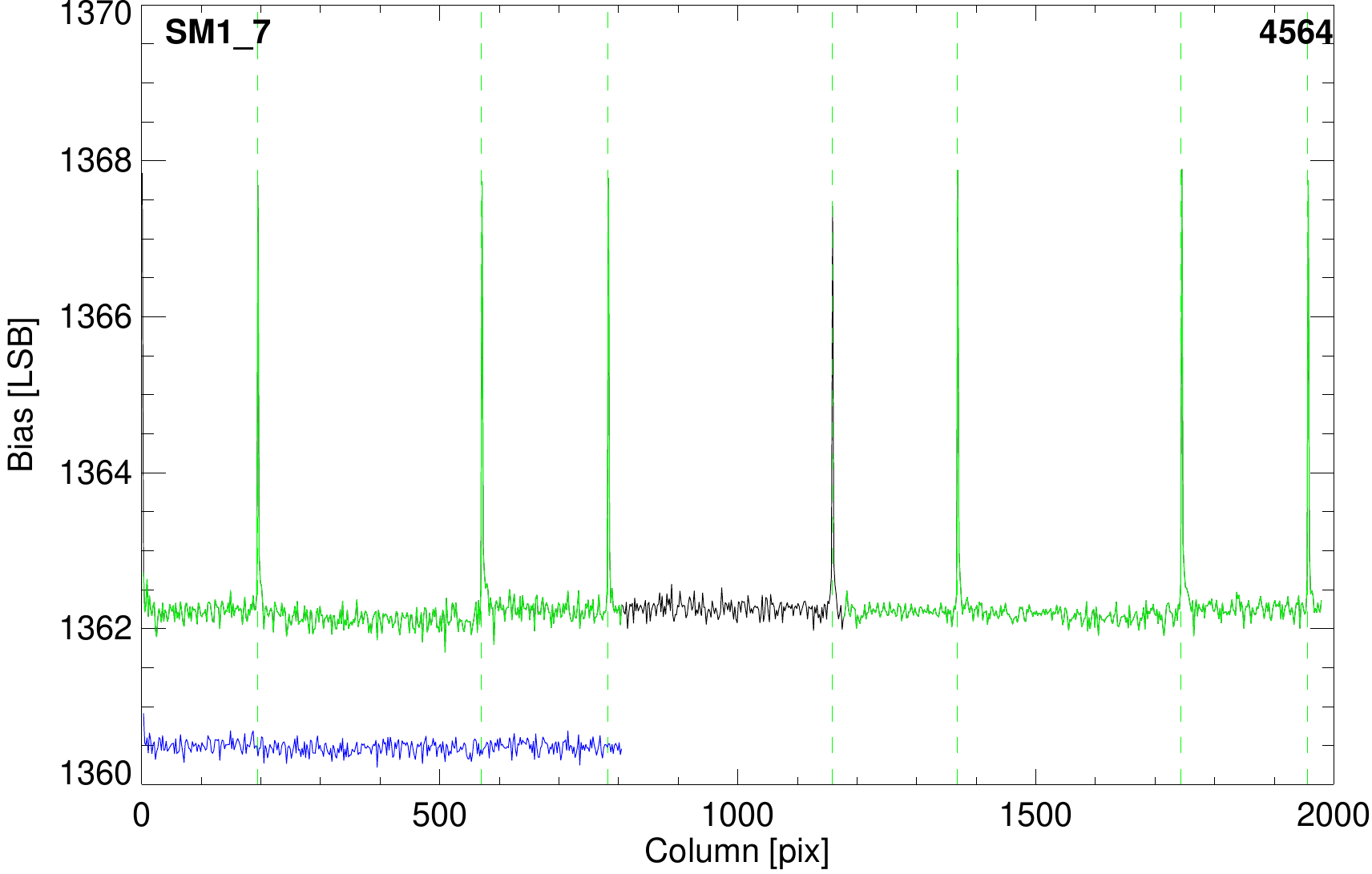}}
\centerline{\includegraphics[scale=0.5]{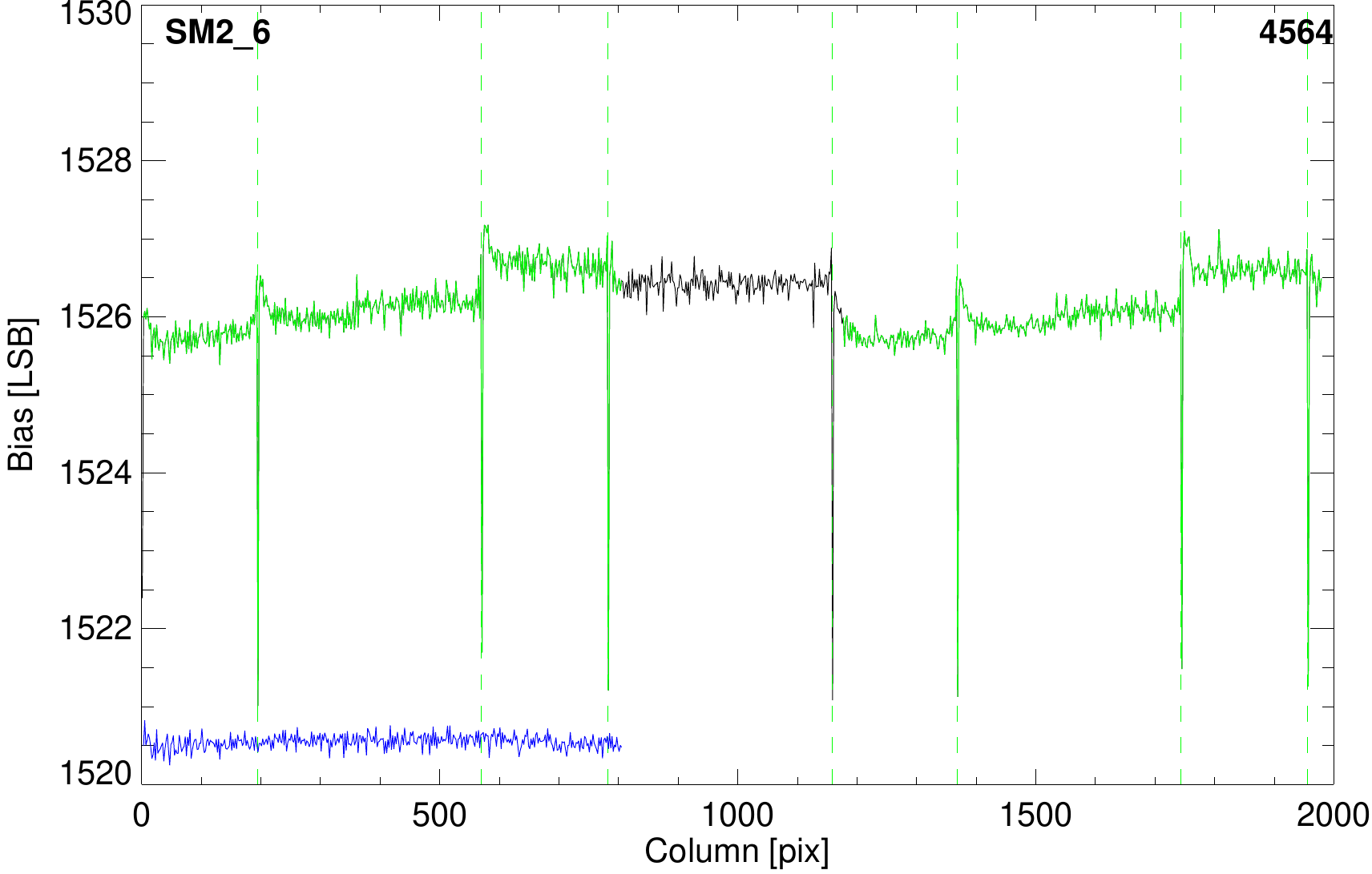}}
\caption[]{Bias for three SM devices (from top to bottom: SM1 in row~4, SM1 in row~7 and SM2 in row~6) from a 
calibration run in December 2016. The two green segments of the bias-curves 
are read in the same phase within a TDI period. The blue curve, with a small
offset, shows the difference between the two green sections divided by square 
root 2. The vertical dashed green lines indicate the readout freezes.
\label{fig:smbias}}
\end{figure}

The sky mapper CCDs are read in full-frame mode, rather than window mode, and 
as such exhibit low--level phase--dependent offset excursions relatively clearly. In
order to have more time for reading a sample and in order to reduce readout
noise, SM CCDs are read with a $2\times2$ pixel binning.  The bias
calibration for these devices is in principle simple, as all that is needed is a set
of dark frames with close to zero exposure time. This can be obtained in
calibration runs with the shortest gate activated, that is,~an exposure time of
just 2\,ms. A detailed calibration model for the readout of SM is therefore not
needed, and because of the along--scan binning by 2 pixels, two TDI periods are available for reading.
Taking the various freeze periods into account, this means that the first 403
samples (covering 806 pixel columns) are read in the same phase within a TDI
period as the last 403 samples. These two intervals are therefore affected by
the same phase--dependent features. Figure~\ref{fig:smbias} shows bias
measurements from December~2016 for three devices. They are based on 3300
consecutive full-frame readings.  The black and green curves show the bias,
where green is used for the two intervals in phase. As the readout covers two
TDI periods, we see seven glitches following readout freezes (vertical dashed
lines). In addition to the rather large glitches, sometimes recovering to a changed
level, we can also see many small spikes, less pronounced in SM1 in row~7, that in
principle could be the result of column defects. This is not the case, however. The
blue curves show the differences between the two green segments, with a
0.5~ADU offset from the bottom of the panels. The differences have been
divided by $\sqrt 2$, such that they would show the same scatter as the green
segments if these were uncorrelated. As the blue curves are much smoother than
the green segments, we can conclude that the main part of the small spikes,
and of the general noisy appearance, is the result of electronic disturbances and not
of defects in the CCD chips. All devices and modes are
subject to such low--level phase--dependent perturbations, and we label this
instability component the intra--TDI phase anomaly, or ITPA.

\begin{figure}
\centerline{\includegraphics[scale=0.4,clip,trim=0 30 0 25]{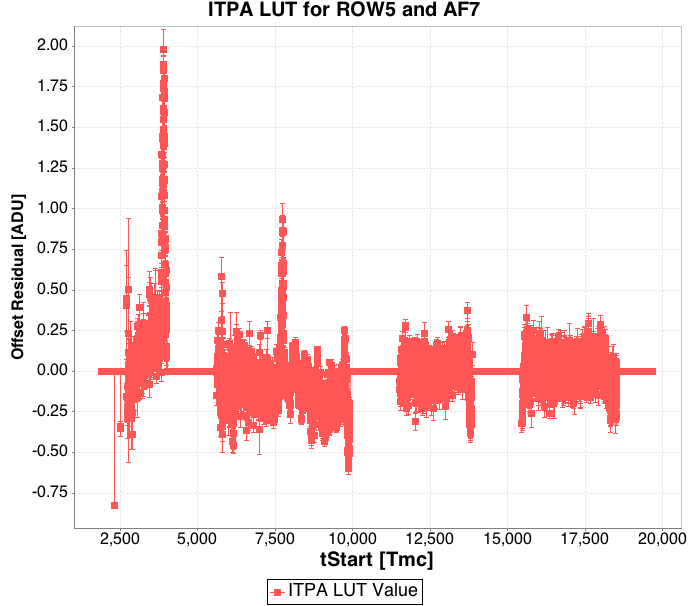}}
\caption[]{Typical example of the intra--TDI phase anomaly observed for device AF7 in row~5 of the \gaia\ focal plane
array during the September~2016 calibration campaign. Sample offsets are plotted as a function of sample start time
in the serial scan and have been corrected for all other components (gross prescan level, common baseline, glitch, and
flush anomalies). The strong features at 3900~Tmc and 7700~Tmc are observed in many devices with amplitudes that are highest
for those nearest the centre of the array, suggesting an electromagnetic pick--up origin.\label{fig:itpaexample}}
\end{figure}

The ITPA morphology is quite different between the various devices and their operating modes
and is not amenable to any simple parametric model. Hence we treat this component as a residual correction after the
previous three have been removed and simply create a look--up table (LUT) of values as a function of sample start time.
Figure~\ref{fig:itpaexample} shows a typical example for AF7 in row~5. 

\subsubsection{Gate--mode-dependent effects}
\label{sec:gatemodeeffects}

\begin{figure}
\centerline{\includegraphics[scale=0.3,clip,trim=65 100 10 22]{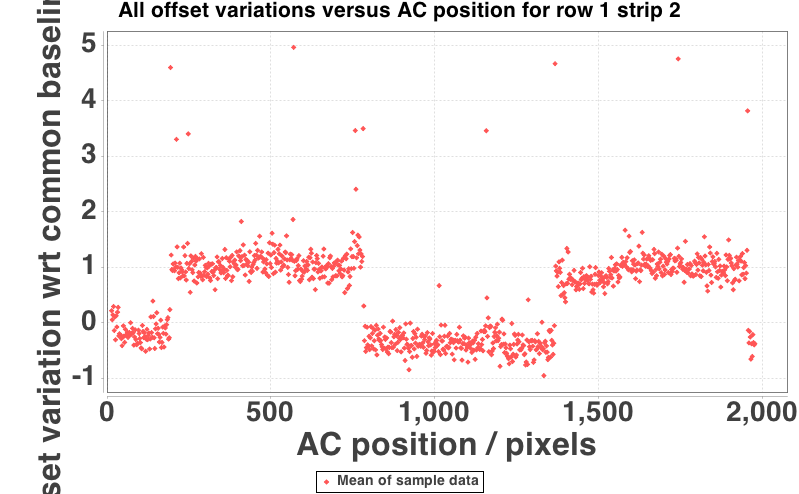}}
\centerline{\includegraphics[scale=0.3,clip,trim=50 25 0 22]{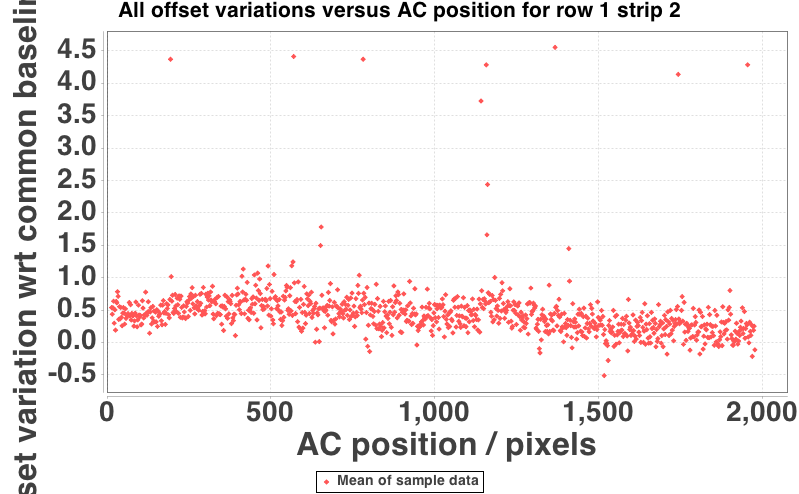}}
\caption[]{Offset morphology in device SM1 in row~1 with one gate (upper panel) and all gates
(lower panel) permanently active during calibration (y-axis units are ADU in both panels).
\label{fig:smgatediffs}}
\end{figure}

During calibration, one or more TDI gates \citep{crowley16} are permanently raised to prevent
photoelectric contamination of the electronic offset measurements (\secref{sec:methods}). It is
inevitable, however, that activating gates in this way will perturb the offset being measured. This results
in measurement of baseline offsets that are not necessarily the same as those applicable during science 
observation modes where gates are only occasionally and transiently active.
The effect is again best illustrated via the offset morphology in SM (full--frame) readout mode. 
Figure~\ref{fig:smgatediffs} shows how the characteristics change between two different gate configurations. 
In particular, the baseline recovery level between each phase clock swing is markedly different between
the two configurations chosen. It is assumed that this effect is present in all devices and modes at
some level, and it explains the zeropoint residual offset seen in, for example, \figref{fig:externalEfficacyHeatMapMedian}.

The strategy chosen to deal with this is to limit permanent gate
activation during calibration, but unfortunately, a handful of astrometric and photometric devices, along
with all RVS devices, require all gates to be raised to obtain a clean calibration. This is because
photoelectric charge builds up behind the TDI gates and can spill over depending on the effectiveness of the gate potential
barrier and the level of stray light that happens to be present during the calibration run. Activating all
gates improves the efficiency of charge dumping (into the lateral anti--blooming drain; \citealt{crowley16})
at as many places as possible across the CCD image area. The non--RVS devices
calibrated in this way are AF3, AF8, and~RP in row~1, AF6 in row~3, and AF5 and~BP in row~4. All other non--RVS
devices have the single gate nearest the serial register permanently active during calibration. Then,
any residual effect resulting from the calibration gate activation mode is dealt with as part of the
baseline offset calibration and takes the form of a correction for these offsets as
determined from the change in prescan level during the calibration period (under the permanent gate 
activation mode) compared with that outside the calibration period. 
There is evidence that the  bias shift resulting from gate activations is not constant over the line readout, 
but can depend upon during which parallel phase that the pixel is read. The possibility of including this in the calibration model for 
future data releases is being examined.

\subsection{Temporal stability}

\begin{figure}
\centerline{\includegraphics[scale=0.25,clip,trim=0 0 0 25]{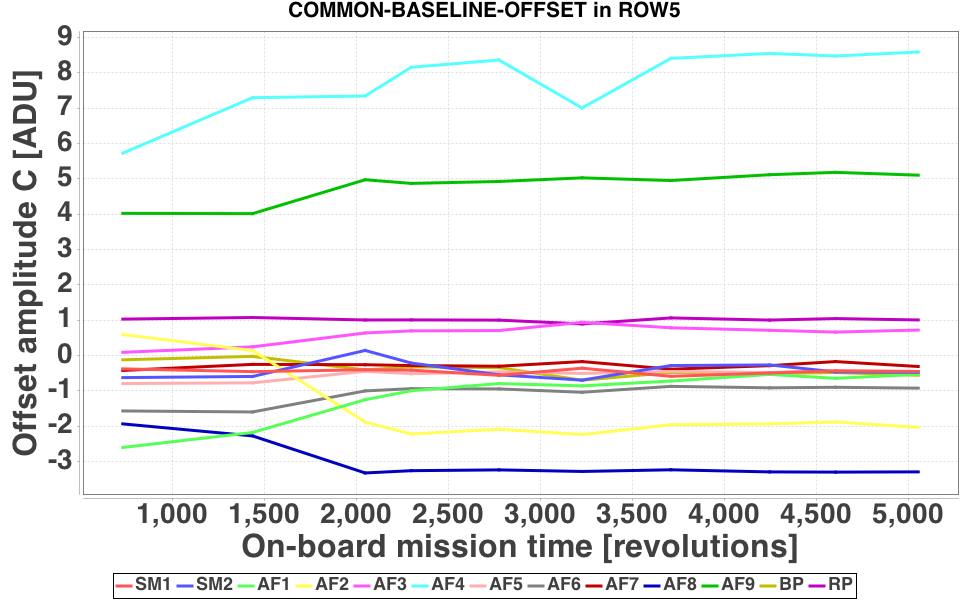}}
\caption[]{Temporal stability of the common baseline offset in SM, AF, and XP devices in row~5. The horizontal
axis is labelled in units of on--board mission time revolutions of \gaia\ (i.e.~units of 6~hours) since just before
the start of science operations in July~2014 (OBMT revolution 980). 
\label{fig:astrophotostability}} 
\end{figure}

\begin{figure*}
\centerline{\includegraphics[scale=0.55,clip,trim=8 8 8 25]{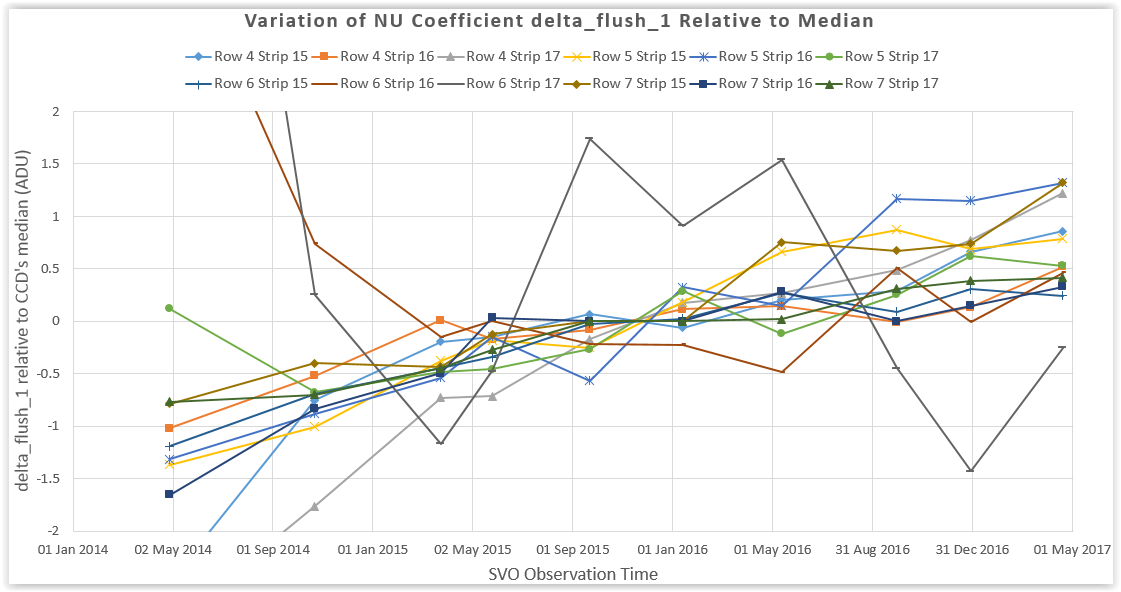}}
\caption[]{Temporal stability of the flush parameter $\Delta_{\rm flush,1}$ in the~12 RVS devices relative
to the median value for each device as measured in all special calibration VO (SVO) datasets 
to date. \label{fig:delta_flush_1_zoom}} 
\end{figure*}

Figure~\ref{fig:astrophotostability} shows the stability of the common baseline offset as a function of time
for the SM, AF, and~XP devices in row~5.
This row of devices exhibits the widest range of variation, but is otherwise typical of the 
astrometric and photometric devices on the entire focal
plane. For the RVS devices, only the flush parameter $\Delta_{\rm flush,1}$ shows any significant variation
with time. This is illustrated in \figref{fig:delta_flush_1_zoom}), which with one exception 
(RVS3 in row~6, strip~17) shows a slow and steady drift in the offset excursion.
The recalibration timescale of  three to four~months, or around~500 revolutions, is sufficient to follow the
observed drifts. 

\section{Discussion}
\label{sec:discussion}

\subsection{Electronic origin}
\label{sec:electronicorigin}

Constraints imposed by the TDI mode of operation  can result in significantly
different conditions on the chip during the readout of a line of pixels. The 
electronic offset is affected by the interplay of applied voltages and voltage
swings over the device, and various shifts and transient effects are visible in a
line readout. A  particularly clear-cut example  are the baseline changes that
are observed between the parallel phase voltage changes that occur multiple
times during the readout of a line. After a phase change, the coupling of the
new voltages applied to the clock lines results in a new baseline offset that is
dependent on the states of the parallel clocking lines during read time.  This is not an effect of the actual clock swings themselves 
(cf.~clock--induced, or spurious charge, e.g.~\citealt{janesick}
and~\citealt{2011MNRAS.411..211T}) but of the
interplay of the new set of voltages that are  applied to the device after the
voltage swings.

As discussed in \secref{sec:results}, the change of state of  one
TDI gate clock line  on the device can indeed noticeably alter the offset level. A
change in offset is also observed when the charge injection pulse is applied
during a pixel readout, but this only manifests itself in some calibration
data acquisitions since during nominal data acquisition, these pulses do not
coincide with a readout.  It is thus apparent that the necessity of interleaving
the pixel reading and parallel clocking implies changes of state to the CCD
that explain some of the more straight-forward systematic disturbances to the
nominal offset.

In addition to the rather straight-forward cross-talk effects described above,
the sky-dependent readout window assignment that is employed on most devices
means that the timing and duration of the pixel-flush sequences is not repeated
line-to-line. As described in \secref{sec:methods}, the flushing of pixels
causes a significant perturbation to the offset on the following sample read, with an
exponential recovery  ($\tau \sim 5~\mu$s). Tests carried out pre-launch using
non-flight electronics showed that  the non-uniformities were still observed on
a test device, demonstrating that the defining characteristics of the effect
originate within the CCD itself.
Two on-chip clock operations that were identified  as being different between
the periods of flushing  and  pixel-reading are that

\begin{enumerate}
\item the reset clock is held constant during the flush sequence,
and \item during flushing, the readout register is clocked at high frequency
(10~MHz) rather than at the operating mode-dependent read rate.
\end{enumerate}

On--ground tests show that the characteristics of the non-uniformities change
dramatically when the duty cycle on both of these clocks during the flush
periods are altered, pointing to coupling paths to the output amplifier
structure.  It is clear that the duty cycle changes to these clocks contribute
significantly to the flush offset phenomenon.

The glitch anomaly (described in \secref{sec:methods}) describes the short-duration spikes ($\tau \sim 100\rm{ns} $) that occur after the resumption of the
serial register clocks. The magnitude of this effect is strongly affected by
changes to the reset pulse. The glitch, despite
occurring on the first pixel read out after the parallel phases changes, is not
a feed-through effect of these clock swings (which are carefully avoided), but
related to the change in duty cycles  on the serial register and reset clocks
during the readout pause. In this respect, the glitch can be thought of as a
similar effect to the settling effect observed on the prescan pixels after line
start.

In addition to the effects mentioned above, for the  fast-readout modes (SM and
AF1), small-scale, often periodic oscillations on  the offset
appear to originate in the PEM itself. Since these features display a rather
fixed pattern in time, they will be removed (along with periodic off-chip
cross-talk signals) as part of the ITPA calibration.

Although not strictly an offset non-uniformity, it is worth pointing out here that
the electronic bias level is affected by the changing of the serial register
clocks between the sampling of the reference level and the video level. The
changes affect the reset level at the output node  and add a fixed electronic
offset to the measured pixel value produced by the correlated double-sampling (CDS) 
operation. This was
confirmed with on--ground testing where the serial clock swing amplitudes were
varied, and another test where the polarity was not changed at all; this effect
is known as the register offset.

In summary, there appears to be a range of different couplings and feed-through
on-chip that combine to affect the pixel offset in a rather complex manner. The
magnitude of the effects that result from flushes and readout pauses might have been reduced by attempting to minimise duty cycle changes and some clock-swing amplitudes. However, the changing of parallel clock states (and gate
clock states, etc.)  that occur during serial readout would still result in some
 unavoidable perturbations on the pixel offsets. For cost and  scheduling
reasons, it was decided not to attempt to modify these operations (some of which
would have required hardware changes to the PEM) and to instead rely on the
on--ground software calibration (in addition to using braking samples on some
CCDs in order to absorb the worst of the flush offset anomaly) since the effects
are deterministic.

\subsection{Possible improvements}

As noted previously, there are several low--level calibration issues that have come to light too late for
correction at DR2 (for which pre-processing was completed more than one year before the time of writing).
These correction enhancements will appear in DR3+, but as other calibrations are refined and as knowledge 
of the payload behaviour improves, it should be possible to do better still. 

Improvements in the acquisition process of the special calibration data are possible. The absence of
a shutter necessitates raising of gates to prevent
photoelectric contamination in the data obtained. Unfortunately, the action of permanently raising gates
disturbs the offsets being measured, resulting in small residuals when applying these corrections to
science data obtained with (generally) no gate active. It will always be necessary to employ at least the gate
nearest the serial register during special calibration runs, but it would be advantageous to limit 
permanent gating to this alone. However, the high levels of stray light create high photoelectric signal
that apparently can build up to levels sufficient to overwhelm single gates, so that for some devices, all gates are
permanently raised during special calibration runs (\secref{sec:gatemodeeffects}). Given such circumstances,
it may be possible to schedule the calibration at spin phase corresponding to minimal stray light. This
may then enable fewer gates to be required to dump the stray photoelectric signal. At the very least, it would generally
minimise the possibility of contamination. Such scheduling is not straight-forward, however, since the
stray light has components resulting from scattered solar, zodiacal, bright solar system objects and very bright star light.

Otherwise, improvements are confined to the on--ground treatment of the problem. Approximately in order of decreasing
significance, these are listed below.
\begin{enumerate}
  \item Accurate on--ground readout reconstruction:  a complete auxiliary data stream is required to reconstruct
  the on--ground the readout history of each device in the focal plane accurately. Any loss of observation logs
  leads to incorrect serial timing with consequent mismatch between the offset excursion model and the offsets
  actually present in the sample data. For example, this is likely the cause of at least some of the very small
  number of under--corrected samples in XP illustrated in \figref{fig:externalXPbeforeafter}, leaving residuals in the 
  range~$-55 e^-$ to~$-35 e^-$ (i.e.~$-14$ to~$-9$~ADU). In rare cases where such an offset is left as a systematic
  underestimate in sample zeropoint over the full length of an XP window, the integrated photometry could be in error by as much as~2~mag
  for the faintest sources. If overestimated, such systematic errors could result in negative integrated flux and rejection of otherwise
  perfectly good observations. Cyclic reprocessing will recover losses resulting from
  telemetry handling software problems during earlier operational phases of the ground segment, 
  although at some level, the data stream will never be
  perfect. Telemetry packet losses are apparent for example during ground-station outages for bad weather, and
  on--board video processing unit operations are very occasionally, and inevitably, interrupted, resulting in 
  loss of information.
  \item Improved flush model: three devices (RVS3 in row~6, strip~17,
and BP and RP in row~3, strips~13 and~14) exhibit calibration 
  residuals at the level of $\sim1$~ADU for unbraked flush offsets over a small range (100 to~200) of flushed
  pixels. This is the range where the onset of the flush excursion changes most rapidly and departs from the simple
  exponential model (\equref{eqn:flush}). A two--component exponential model was elaborated during initial 
  on--ground investigations (\secref{sec:intro}), but has not been implemented so far as the final stages of
  limited testing prior to launch indicated that this was an unnecessary complication.
  \item Better handling of flush sequences interrupted by readout freezes: there is some 
  evidence~\citep{Boudreault:12} that the model serial dependency for tracking the evolution of the offset
  excursion is not perfect for all devices in all circumstances. In particular, the (very rare) case when a 
  long flush sequence is interrupted by two freezes has not been studied in any detail.
  \item Parallel phase--dependent gate mode offsets: as discussed in \secref{sec:electronicorigin},
  a difference in offset between the first/fourth and second/third parallel phase times is possible for
  some devices depending on electronic coupling. At present, a single offset delta is measured between the calibration
  gate mode and normal science operations mode via the prescan level inside and outside the special calibration
  period and is applied regardless of parallel phase zone in the serial scan. This means that any phase--dependent
  difference resulting from the gate activations present during calibration will not be applied, and in any case
  may be inappropriate for science mode operations. 
\end{enumerate}
This last item is particularly problematic because i) there is no way of calibrating the offset non--uniformities
in a single, separable process without activating at least the gate nearest the serial register; and ii) while for the most part, no gate, or at most one gate, is active during sample readout in science-mode operations, in
principle any one or more gates out of the~eight employed could be active. The latter results in
potentially~256 different gate combinations per device, requiring offset calibration. The former requires
that calibration be done from the science data rather than as a separable process based on special calibration 
sequences. Any calibration process that is based on the science data alone faces the awkward problem of
entanglement and non--independence in that all other calibrations have to be assumed perfect in order that residual
patterns observed in some subset of the data be attributable to any one process. For example, data such as
those present in \figref{fig:externalAFbeforeafter} have been corrected for thermo-- and photoelectric 
background. This correction was minimised by limiting the dataset to sample integrations of 16~ms (again taking
advantage of gate activations affecting these samples). Further subdividing this already limited dataset in order to
tease out hundreds of calibration parameters is challenging; relaxing the integration time limit
to pull in more calibration data reduces the independence from the multiplicative calibrations (dark and background
signal). Despite these complexities, it is likely that the bias calibration residuals can ultimately be reduced
everywhere below the~0.5~ADU level only by employing some subset of the science data in each device.
In any case, the residuals are already reduced to the level where they have an insignificant impact on the video-chain total detection noise budget.

\section{Conclusion}
\label{sec:conclusion}

We have presented a detailed study of the behaviour of the electronic offsets in the \gaia\ CCD video chains.
We have described how issues concerning the stability of the digitisation zeropoints were identified 
during on--ground testing. While hardware mitigation schemes were insufficient to fully stabilise the offsets,
we have described how laboratory testing showed the remaining effects to be deterministic, and hence amenable
to treatment in on--ground processing.
During preparations prior to launch, a complete mitigation strategy was put in place, including a calibration
mechanism and the development of pipeline processing software to correct the so--called bias non--uniformities.
We have demonstrated the in--orbit behaviour of the offsets, showing them to behave in the ways anticipated
before launch. We have illustrated the calibration and mitigation process and have shown how the fundamental
video-chain detection noise limit is recovered in science data corrected for the offset instabilities in
the vast majority of samples forming the basis of Data Release~2. We have briefly discussed low--level
effects that remain at this point in the cyclic data processing and that will nonetheless be corrected in
the future data releases.

%

\begin{acknowledgements}

We also wish to acknowledge the role of Airbus Defence \& Space (ADS) for the CCD-PEM detector system, 
the FPA and supplying on--ground test data to the consortium prior to launch of \gaia. The initial
bias NU calibration model and the initial special calibration sequences were all designed by ADS.

This work has made use of results from the ESA space mission \gaia, the data
from which were processed by the \gaia\ Data Processing and Analysis Consortium
(DPAC). Funding for the DPAC has been provided by national institutions, in
particular the institutions participating in the \gaia\ Multilateral Agreement.
The \gaia\ mission website is \url{http://www.cosmos.esa.int/gaia}.

Many of the authors are members of the \gaia\ Data Processing and Analysis Consortium
(DPAC), and their work has been supported by the following funding agencies: 
the United Kingdom Science and Technology Facilities Council (\gaia\ Data Flow System
grants); the United Kingdom Space Agency (\gaia\ Post Launch Support grant); and
MINECO (Spanish Ministry of Economy) through grants ESP2016-80079-C2-1-R (MINECO/FEDER, UE),
ESP2014-55996-C2-1-R (MINECO/FEDER, UE) and MDM-2014-0369 of ICCUB (Unidad de Excelencia
'Mar\'ia de Maeztu').

Finally, we thank the anonymous referee for a positive reception of this rather technical
paper and for several useful suggestions that have improved the clarity of presentation.

\end{acknowledgements}

\bibliographystyle{aa} 
\bibliography{refs} 


\end{document}